\documentclass[aps,prb,twocolumn,showpacs,groupedaddress]{revtex4}

\usepackage{graphicx}
\usepackage{amssymb}

\begin{document}


\title{Dynamic scaling and aging phenomena in short-range Ising spin 
glass: Cu$_{0.5}$Co$_{0.5}$Cl$_{2}$-FeCl$_{3}$ graphite 
bi-intercalation compound}


\author{Itsuko S. Suzuki }
\email[]{itsuko@binghamton.edu}
\affiliation{Department of Physics, State University of New York at 
Binghamton, Binghamton, New York 13902-6016}

\author{Masatsugu Suzuki }
\email[]{suzuki@binghamton.edu}
\affiliation{Department of Physics, State University of New York at 
Binghamton, Binghamton, New York 13902-6016}


\date{\today}

\begin{abstract}
Static and dynamic behavior of short-range Ising-spin glass
Cu$_{0.5}$Co$_{0.5}$Cl$_{2}$-FeCl$_{3}$ graphite bi-intercalation compounds
(GBIC) has been studied with SQUID DC and AC magnetic susceptibility.  The
$T$ dependence of the zero-field relaxation time $\tau$ above a
spin-freezing temperature $T_{g}$ (= 3.92 $\pm$ 0.11 K) is well described
by critical slowing down.  
The absorption $\chi^{\prime\prime}$ below $T_{g}$ decreases with
increasing angular frequency $\omega$, which is in contrast to the case of
3D Ising spin glass.
The dynamic freezing temperature
$T_{f}(H,\omega)$ at which d$M_{FC}(T,H)/$d$H=\chi^{\prime}(T,H=0,\omega)$,
is determined as a function of frequency (0.01 Hz $\leq \omega/2\pi \leq$ 1
kHz) and magnetic field (0 $\leq H \leq$ 5 kOe).  The dynamic scaling
analysis of the relaxation time $\tau(T,H)$ defined as $\tau = 1/\omega$ at
$T = T_{f}(H,\omega)$ suggests the absence of SG phase in the presence of
$H$ (at least above 100 Oe).  Dynamic scaling analysis of $\chi^{\prime
\prime}(T, \omega)$ and $\tau(T,H)$ near $T_{g}$ leads to the critical
exponents ($\beta$ = 0.36 $\pm$ 0.03, $\gamma$
= 3.5 $\pm$ 0.4, $\nu$ = 1.4 $\pm$ 0.2, $z$ =
6.6 $\pm$ 1.2, $\psi$ = 0.24 $\pm$
0.02, and $\theta$ = 0.13 $\pm$ 0.02).  The
aging phenomenon is studied through the absorption $\chi^{\prime
\prime}(\omega, t)$ below $T_{g}$.  It obeys a $(\omega t)^{-b^{\prime
\prime}}$ power-law decay with an exponent $b^{\prime \prime}\approx 0.15 -
0.2$.
The rejuvenation effect is also observed under sufficiently large 
(temperature and magnetic-field) perturbations.
\end{abstract}

\pacs{75.50.Lk, 75.40.Gb, 75.40.Cx}

\maketitle



\section{\label{intro}Introduction}
In recent years there has been much effort to understand phase transitions
of three-dimensional (3D) Ising spin glasses (SG's) with short-range
interactions.  The dynamic behavior of the low temperature SG phase has
been a subject of some controversy.  There are basically two different
pictures of the SG phase, the mean-field picture (including Monte Carlo
simulations)\cite{Marinari1998} and the scaling
picture.\cite{McMillan1984,Fisher1986,Fisher1988a,Fisher1988b,Bray1987,Bray1988}

The mean-field picture is originated from the replica-symmetry-breaking
(RSB) solution of Parisi for the infinite-range Sherrington-Kirkpatrick
(SK) model.\cite{Marinari1998} In this picture the SG phase has a
complicated free energy structure with a number of degenerate minimum
separated with each other by large free energy barriers.  The SG phase
survives in the presence of a magnetic field.  A de Almeida-Thouless (AT)
line separates the SG phase from a paramagnetic phase.\cite{Almeida1978}

The scaling picture (so-called the droplet model) is based on
renormalization group arguments.
\cite{McMillan1984,Fisher1986,Fisher1988a,Fisher1988b,Bray1987,Bray1988}
In this picture there are only two thermodynamic states related to each
other by a global spin flip, and the important excitations at low
temperatures are droplets of overturned spins.  In three dimensions a true
phase transition should exist only in zero field and no irreversibility
should be seen in the presence of any finite magnetic field.

Because of extremely long relaxation time, there is no direct measurement
available that can demonstrate whether the SG phase exists even in zero
magnetic field.\cite{Mattsson1995,Nordblad1998,Petit1999} The dynamic
scaling analysis is required to overcome such an experimental difficulty in
confirming the existence of thermally equilibrium SG phase in a finite
magnetic field.  This problem has been addressed by Mattsson et
al.\cite{Mattsson1995} They have studied the dynamic susceptibility of a 3D
Ising SG, Fe$_{0.5}$Mn$_{0.5}$TiO$_{3}$.  The in-field scaling behavior of
the dynamic susceptibility, which is remarkably different from the
zero-field behavior, is explained as demonstrating the absence of SG phase
in a finite field.

In this paper we study the slow dynamics of the SG phase in a short-range
3D Ising SG, Cu$_{0.5}$Co$_{0.5}$Cl$_{2}$-FeCl$_{3}$ graphite
bi-intercalation compound (GBIC), using SQUID DC and AC magnetic
susceptibility.  This compound undergoes a SG transition at a spin freezing
temperature $T_{g}$ (= 3.92 $\pm$ 0.11 K) at zero magnetic field. 
Following a method used by Mattsson et al.,\cite{Mattsson1995} we determine
the dynamic freezing temperature $T_{f}(H,\omega)$ at which d$M_{FC}(T,H)/$d$H
= \chi^{\prime}(T,H=0,\omega)$, as a function of frequency (0.01 Hz $\leq
\omega/2\pi \leq$ 1 kHz) and field (0 $\leq H \leq$ 5 kOe).  The dynamic
scaling analysis of the relaxation time $\tau(T,H)$ defined as $\tau =
1/\omega$ at $T = T_{f}(H,\omega)$, suggests that the SG phase does not
survive in the presence of $H$ (at least above 100 Oe).  The critical
exponents are determined from the dynamic scaling analysis for
$\chi^{\prime \prime}(T,\omega)$ and $\tau(T,H)$ and compared to those for
Fe$_{0.5}$Mn$_{0.5}$TiO$_{3}$.  The aging phenomenon is also studied
through the time dependence of the absorption $\chi^{\prime
\prime}(\omega,t)$ below $T_{g}$.  It obeys a ($\omega t)^{-b^{\prime
\prime}}$ power-law decay with an exponent $b^{\prime \prime}$ $\approx$
0.15 $-$ 0.2.

Cu$_{0.5}$Co$_{0.5}$Cl$_{2}$-FeCl$_{3}$ GBIC has a unique layered structure
where the Cu$_{0.5}$Co$_{0.5}$Cl$_{2}$ intercalate layer (= $I_{1}$) and
FeCl$_{3}$ intercalate layers (=$I_{2}$) alternate with a single graphite
layer ($G$), forming a stacking sequence
(-$G$-$I_{1}$-$G$-$I_{2}$-$G$-$I_{1}$-$G$-$I_{2}$-$G$-$\cdot$) along the
$c$ axis.\cite{Suzuki1997,Suzuki1999a} In the Cu$_{0.5}$Co$_{0.5}$Cl$_{2}$
intercalate layer two kinds of magnetic ions (Cu$^{2+}$ and Co$^{2+}$) are
randomly distributed on the triangular lattice.  The spin order in the
Cu$_{0.5}$Co$_{0.5}$Cl$_{2}$ layers is coupled with that in the FeCl$_{3}$
intercalate layer through an interplanar exchange interaction,
leading to the spin frustration effect.

The XY and Ising character of the present system are due to the easy-plane
type anisotropy field $H_{A}^{out}$ and the in-plane anisotropy field
$H_{A}^{in}$, respectively: $H_{A}^{out} \gg H_{A}^{in}$.  Although the
in-plane structure of Cu$_{0.5}$Co$_{0.5}$Cl$_{2}$ layers is incommensurate
with that of graphene sheets, Cu and Co atoms tend to sit over hexagon
centers of graphene sheets because of the interplanar structural 
correlation.  The sixfold symmetry of graphen sheets imposes the in-plane
spin anisotropy in the $c$ plane.  Because of this, the spins tend to align
along the easy axis with sixfold symmetry in the $c$ plane.\cite{Enoki2003}

It is well known that the intercalate layers are formed of small 
islands in acceptor graphite intercalation compounds (GIC's).\cite{Enoki2003}
In stage-2 CoCl$_{2}$ GIC, for example, there is a charge transfer 
from graphene sheets to the CoCl$_{2}$ intercalate layer. The 
periphery 
of such islands provides sites for charges transferred. The size 
of islands is on the order of 400 {\AA}. In spite of few structural 
studies, it is predicted that similar island structures exist 
both in the Cu$_{0.5}$Co$_{0.5}$Cl$_{2}$ and FeCl$_{3}$ layers of our system.

\section{\label{exp}EXPERIMENTAL PROCEDURE}
A sample of stage-2 Cu$_{0.5}$Co$_{0.5}$Cl$_{2}$ graphite intercalation
compound (GIC) as a starting material was prepared from single crystal kish
graphite (SCKG) by vapor reaction of anhydrated
Cu$_{0.5}$Co$_{0.5}$Cl$_{2}$ in a chlorine atmosphere with a gas pressure
of $\approx$ 740 Torr.\cite{Suzuki1998,Suzuki1999b} The reaction was
continued at 500 $^\circ$C for three weeks.  The sample of
Cu$_{0.5}$Co$_{0.5}$Cl$_{2}$-FeCl$_{3}$ GBIC was prepared by a sequential
intercalation method: the intercalant FeCl$_{3}$ was intercalated into
empty graphite galleries of stage-2 Cu$_{0.5}$Co$_{0.5}$Cl$_{2}$ GIC. A
mixture of well-defined stage-2 Cu$_{0.5}$Co$_{0.5}$Cl$_{2}$ GIC based on
SCKG and single-crystal FeCl$_{3}$ was sealed in vacuum inside Pyrex glass
tubing, and was kept at 330 $^\circ$C for two weeks.  The stoichiometry of
the sample is represented by
C$_{m}$(Cu$_{0.5}$Co$_{0.5}$Cl$_{2}$)$_{1-c}$(FeCl$_{3}$)$_{c}$.
The concentration of C and Fe ($m$ and $c$) was determined from weight
uptake measurement and electron microprobe measurement [using a scanning
electron microscope (Model Hitachi S-450)]: $m$ = 5.26 $\pm$ 0.05 and $c$ =
0.53 $\pm$ 0.03.  The (00$L$) x-ray diffraction measurements of stage-2
Cu$_{0.5}$Co$_{0.5}$Cl$_{2}$ GIC and
Cu$_{0.5}$Co$_{0.5}$Cl$_{2}$-FeCl$_{3}$ GBIC were made at 300 K by using a
Huber double circle diffractometer with a MoK$\alpha$ x-ray radiation
source (1.5 kW).  The $c$-axis repeat distance of stage-2
Cu$_{0.5}$Co$_{0.5}$Cl$_{2}$ GIC and
Cu$_{0.5}$Co$_{0.5}$Cl$_{2}$-FeCl$_{3}$ GBIC was determined as 12.83 $\pm$
0.05 $\AA$ and 18.81 $\pm$ 0.05 $\AA$, respectively.

The DC magnetization and AC susceptibility were measured using a SQUID
magnetometer (Quantum Design, MPMS XL-5) with an ultra low field capability
option.  First a remnant magnetic field was reduced to zero field (exactly
less than 3 mOe) at 298 K for both DC magnetization and AC susceptibility
measurements.  Then the sample was cooled from 298 K to 1.9 K in a zero
field.  (i) \textit{Measurements of the zero-field cooled susceptibility
($\chi_{ZFC}$) and the field-cooled susceptibility ($\chi_{FC}$).} After an
external magnetic field $H$ (0 $\leq H \leq$ 1 kOe) was applied along the
$c$ plane (basal plane of graphene layer) at 1.9 K, $\chi_{ZFC}$ was
measured with increasing temperature ($T$) from 1.9 to 20 K. After
annealing of sample for 10 minutes at 50 K in the presence of $H$,
$\chi_{FC}$ was measured with decreasing $T$ from 20 to 1.9 K. (ii)
\textit{AC susceptibility measurement}.  The frequency ($f$), magnetic
field, and temperature dependence of the dispersion ($\chi^{\prime}$) and
absorption ($\chi^{\prime \prime}$) was measured between 1.9 to 20 K, where
the frequency of the AC field is $f$ = 0.01 - 1000 Hz and the amplitude $h$
is typically $h$ = 50 mOe.

\section{\label{result}RESULT}
\subsection{\label{resultA}$\chi_{ZFC}(T,H)$ and $\chi_{FC}(T,H)$}

\begin{figure}
\includegraphics[width=8.0cm]{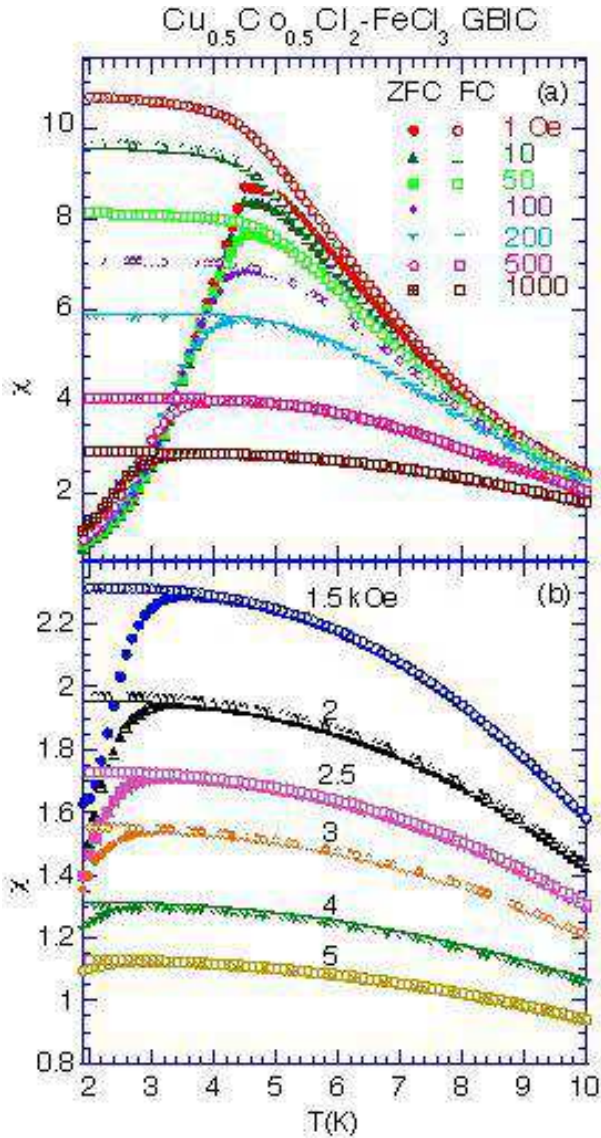}
\caption{\label{fig:one}(a) and (b) $T$ dependence of $\chi_{ZFC}$ and
$\chi_{FC}$ at various $H$ for Cu$_{0.5}$Co$_{0.5}$Cl$_{2}$-FeCl$_{3}$
GBIC. $H$ is applied along the $c$ plane (graphene basal plane)
perpendicular to the $c$ axis.}
\end{figure}

\begin{figure}
\includegraphics[width=8.0cm]{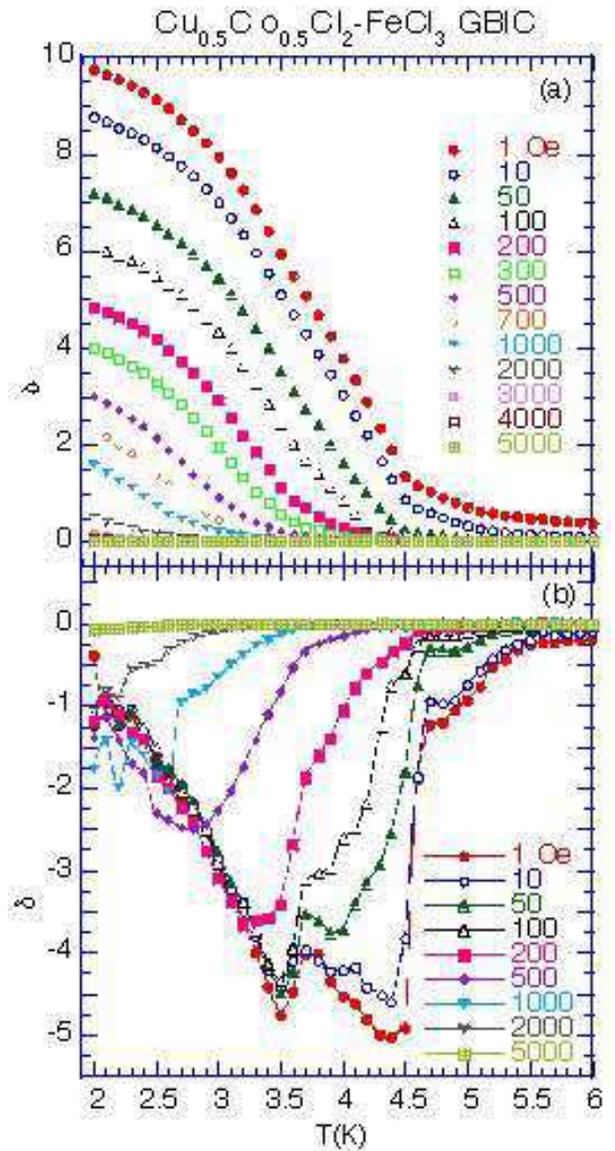}
\caption{\label{fig:two}$T$ dependence of (a) $\delta$ (= $\chi_{FC}
-\chi_{ZFC}$) and (b) d$\delta$/d$T$ at various $H$.  The value of
$\delta$ is derived from Fig.~\ref{fig:one}. The solid lines are 
guides to the eyes.}
\end{figure}

Figures \ref{fig:one}(a) and (b) show the $T$ dependence of the ZFC and FC
susceptibilities ($\chi_{ZFC}$ and $\chi_{FC}$) of
Cu$_{0.5}$Co$_{0.5}$Cl$_{2}$-FeCl$_{3}$ GBIC at various $H$, where the
magnetic field ($H$) is applied along the $c$ plane perpendicular to the
$c$ axis.  The $T$ dependence of $\chi_{ZFC}$ and $\chi_{FC}$ is similar to
that of Fe$_{0.5}$Mn$_{0.5}$TiO$_{3}$.\cite{Ito1986} The susceptibility
$\chi_{ZFC}$ at $H$ = 1 Oe exhibits a cusp around 4.5 K , while $\chi_{FC}$
is nearly temperature independent below 3.9 K.
Note that $\chi_{FC}$ along the $c$ axis at $H$ = 1 
Oe is on the order of one-tenth of $\chi_{FC}$ along the $c$ 
plane around 3.9 K. Figure \ref{fig:two}(a)
shows the $T$ dependence of the difference $\delta$ defined by $\delta
=\chi_{FC} - \chi_{ZFC}$ at various $H$, which is the measure for the
irreversibility of the susceptibility.  The deviation of $\chi_{ZFC}$ from
$\chi_{FC}$ at $H$ = 1 Oe starts to occur below 10 K, and drastically
increases with decreasing $T$ below the cusp temperature of $\chi_{ZFC}$. 
The inflection-point temperature of $\delta$ vs $T$ corresponds to a local
minimum temperature of d$\delta$/d$T$ vs $T$ shown in
Fig.~\ref{fig:two}(b).  The inflection-point temperature of $\delta$
drastically decreases with increasing $H$, leading to the AT line in the
$(H,T)$ diagram (see Sec.~\ref{resultD} for further discussion).
The non-vanishing $\delta$ well above the inflection-point 
temperature is partly due to the islandic nature of Cu$_{0.5}$Co$_{0.5}$Cl$_{2}$ 
and FeCl$_{3}$ layers. The growth of the in-plane spin correlation length is 
partly limited by the existence of small islands, leading to the 
smearing of the inflection-point temperature over the system.

\subsection{\label{resultB}Relaxation time: $\chi^{\prime}(T,\omega)$ and 
$\chi^{\prime \prime}(T,\omega)$}

\begin{figure}
\includegraphics[width=8.0cm]{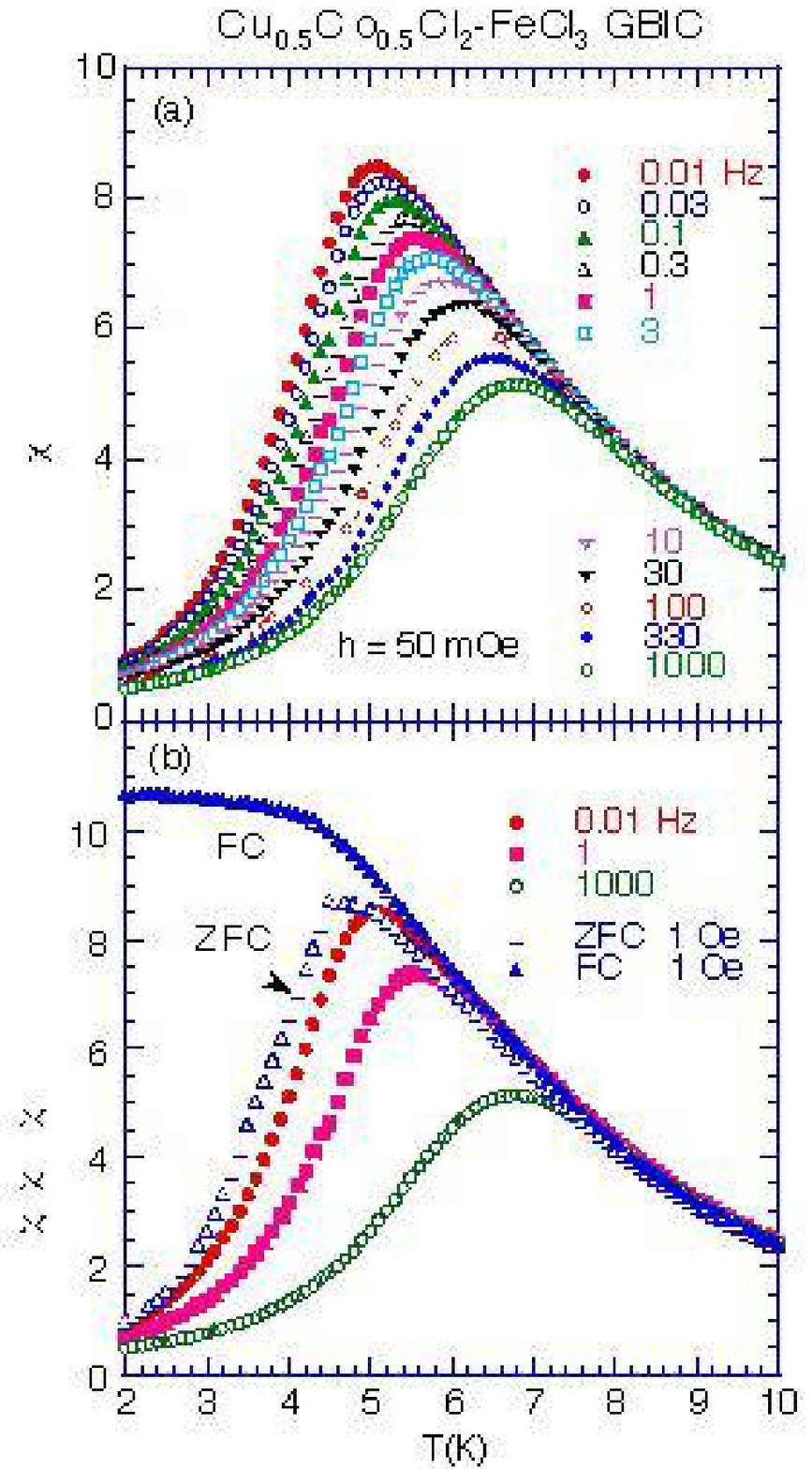}
\caption{\label{fig:three}(a) $T$ dependence of $\chi^{\prime}$ at various
$f$.  $h$ = 50 mOe.  $H$ = 0.  (b) $T$ dependence of $\chi_{ZFC}$ and
$\chi_{FC}$ ($H$ = 1 Oe), which is compared with that of $\chi^{\prime}$ at
$f$ = 0.01, 1 and 1000 Hz ($H$ = 0).}
\end{figure}

\begin{figure}
\includegraphics[width=8.0cm]{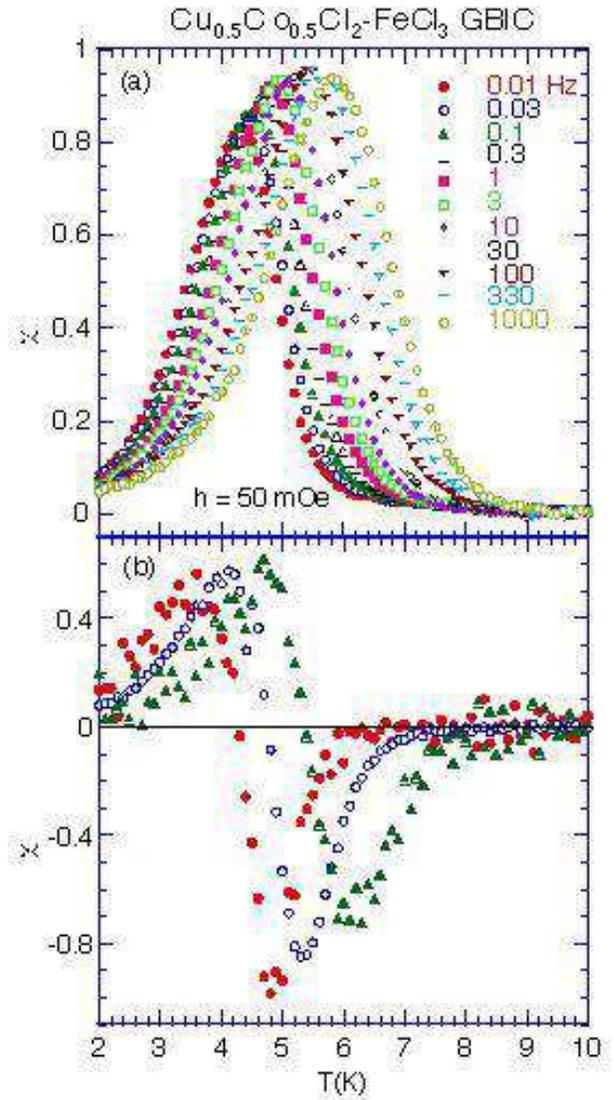}
\caption{\label{fig:four}$T$ dependence of (a) $\chi^{\prime \prime}$ and
(b) d$\chi^{\prime \prime}$/d$T$ at various $f$.  $h$ = 50 mOe.  $H$ = 0.}
\end{figure}

\begin{figure}
\includegraphics[width=8.0cm]{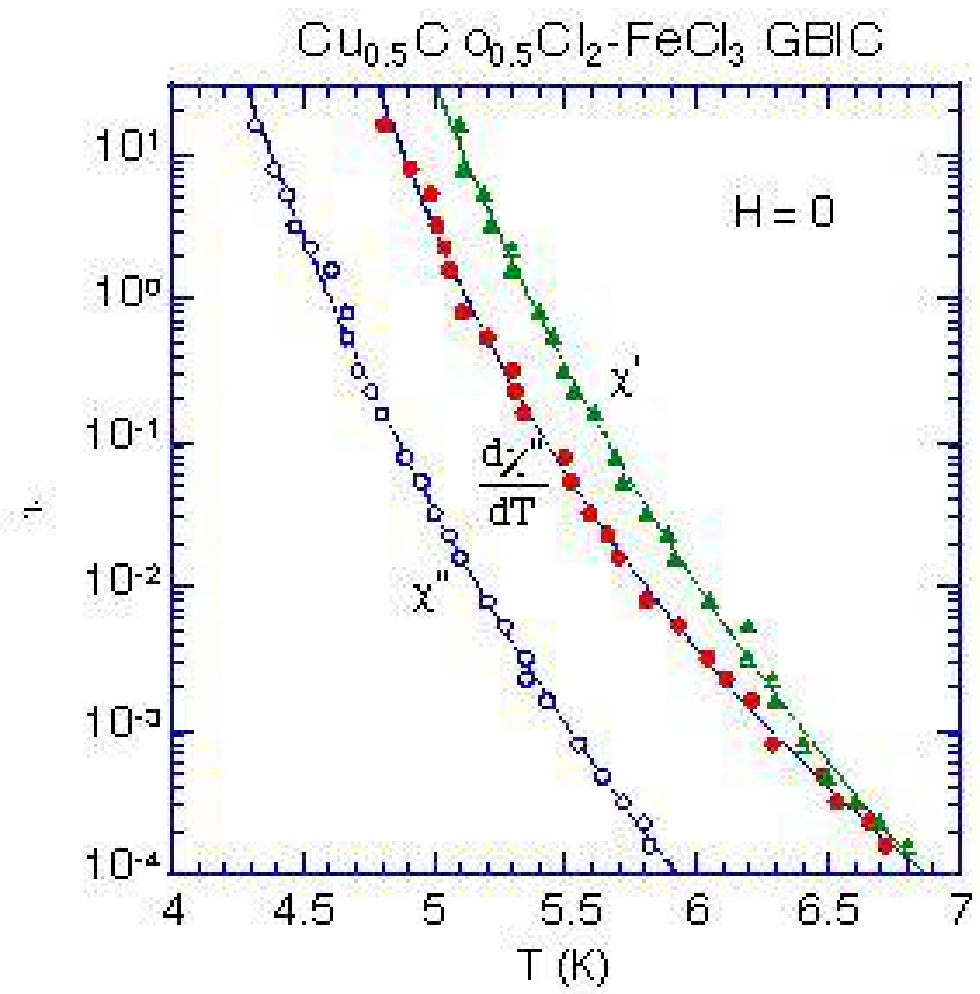}
\caption{\label{fig:five}$T$ dependence of relaxation time $\tau$, which is
determined from peak temperatures of $\chi^{\prime \prime}$ vs $T$ and
$\chi^{\prime}$ vs $T$, and local minimum temperature of d$\chi^{\prime
\prime}$/d$T$ vs $T$.  The solid lines are fits to Eq. (\ref{eq:one})
(critical slowing down) with parameters $x$, $T_{g}$, and $\tau^{*}$
given in the text.}
\end{figure}

In Fig.~\ref{fig:three}(a) we show the $T$ dependence of the dispersion
$\chi^{\prime}(T,\omega)$ at various frequencies ($f$ = 0.01 $-$ 1000 Hz),
where $\omega$ (= 2$\pi f$) is the angular frequency and $h$ (= 50 mOe) is
the amplitude of the AC field.  In Fig.~\ref{fig:three}(b), the $T$
dependence of $\chi_{ZFC}$ and $\chi_{FC}$ at $H$ = 1 Oe is compared to
typical data of $\chi^{\prime}(T,\omega)$.  The susceptibility $\chi_{FC}$
may correspond to the equilibrium susceptibility ($\chi_{eq}$).  In
Fig.~\ref{fig:four}(a) we show the $T$ dependence of the absorption
$\chi^{\prime \prime}(T,\omega)$ at various $f$.  Similar behaviors in
$\chi^{\prime}(T,\omega)$ and $\chi^{\prime \prime}(T,\omega)$ have been
observed in Fe-C nanoparticles by Hansen et al.\cite{Hansen2002} Figure
\ref{fig:four}(b) shows the $T$ dependence of the derivative d$\chi^{\prime
\prime}(T, \omega)$/d$T$.  If we assume conventional critical slowing down
on approaching the SG transition temperature $T_{g}$ from the high-$T$
side, the $T$ dependence of the relaxation time $\tau$ can be described
by\cite{Gunnarsson1988}
\begin{equation} 
\tau = \tau^{*}(T/T_{g} - 1)^{-x},
\label{eq:one} 
\end{equation} 
with $x = \nu z$, where $z$ is the dynamic critical exponent, $\nu$ is the
critical exponent of the spin correlation length $\xi$, and $\tau^{*}$ is
the characteristic time.  In the analysis of $\tau$ vs $T$, $T$ corresponds
to the local minimum temperature of d$\chi^{\prime \prime}$/d$T$ vs $T$, or
the peak temperatures of $\chi^{\prime \prime}$ vs $T$ and $\chi^{\prime}$
vs $T$, where $\tau$ is set equal to $\omega^{-1}$.  In Fig.~\ref{fig:five}
we show the $T$ dependence of $\tau$ thus obtained.  The least squares fits
of the data of $\tau$ vs $T$ yield $x$ = 10.3 $\pm$ 0.7, $T_{g}$ = 3.92
$\pm$ 0.11 K, $\tau^{*} = (5.27 \pm 0.07) \times 10^{-6}$ sec for the local
minimum temperature of d$\chi^{\prime \prime}$/d$T$ vs $T$, $x = 15.5 \pm
1.5$, $T_{g} = 2.98 \pm 0.18$ K, $\tau^{*} = (8.56 \pm 0.05) \times
10^{-5}$ sec for the peak temperature of $\chi^{\prime \prime}$ vs $T$, and
$x = 15.3 \pm 2.0$, $T_{g} = 3.57 \pm 0.27$ K, $\tau^{*} = (3.08 \pm 0.10)
\times 10^{-5}$ sec for the peak temperature of $\chi^{\prime}$ vs $T$. 
The parameters thus obtained are rather different.  The relaxation time
$\tau^{*}$ is much larger than a microscopic relaxation time $\tau_{0}$
(typically 10$^{-10}$ sec).  Here we assume that the local minimum
temperature of d$\chi^{\prime \prime}$/d$T$ vs $T$ at $\omega$ = 0
corresponds to the spin freezing temperature $T_{g}$.  $T_{g}$ is close to
a temperature below which $\chi_{FC}$ at $H$ = 1 Oe becomes nearly
constant.  The value of $x$ (= 10.3 $\pm$ 0.7) is rather close to that
predicted by Ogielski for the 3D $\pm J$ Ising SG ($x = 7.9 \pm
1.0$)\cite{Ogielski1985} from Monte Carlo (MC) simulations.  The situation
is a little different for
Fe$_{0.5}$Mn$_{0.5}$TiO$_{3}$.\cite{Gunnarsson1988} In the analysis of
$\tau$ vs $T$, $T$ is determined either as the maximum of $\chi^{\prime}$
or as the inflection point of $\chi^{\prime \prime}$.  In both cases,
$\tau$ is well described by Eq. (\ref{eq:one}) with $T_{g} = 20.95 
\pm 0.1$ K, $x
= 10.0 \pm 1.0$, and log$_{10}\tau_{0} = -12.8 \pm 1.0$, where $\tau_{0}$
is a microscopic relaxation time.  The value of $x$ is in excellent
agreement with our result.

Here we assume a generalized form of the relaxation time, more suitable to
the description of SG behavior near $T_{g}$.\cite{Bouchaud2001} The
relaxation time $\tau(l,T)$, which is needed to flip the $l$-sized cluster
of spins, is governed by thermal activation over a barrier $B_{T}(l)$, in
such a way that
\begin{equation} 
\tau(l,T) = \tau_{0}l^{z} \exp [B_{T}(l)/k_{B}T],
\label{eq:two} 
\end{equation} 
where $l$ is in units of the lattice constant $a$.  The energy barrier is
described by $B_{T}(l) =Y(T)l^{\psi}$, where $\psi$ is the barrier
exponent.  The wall stiffness $Y(T)$ should vanish above $T_{g}$ like in
ferromagnets and is described by $Y(T) =Y_{0}(1-T/T_{g})^{\psi \nu}$ below
$T_{g}$, where $\nu$ is the critical exponent of spin correlation length
$\xi$ and $Y_{0}$ is on the order of $k_{B}T_{g}$.  Above $T_{g}$ the
exponential term in Eq.(\ref{eq:two}) is nearly equal to 1, leading to the
expression for the critical slowing down above $T_{g}$: $\tau_{+}(l,T)
\approx \tau_{0}l^{z}$.  

\subsection{\label{resultC}Dynamic scaling of $T\chi^{\prime \prime}(T,\omega)$}

\begin{figure}
\includegraphics[width=8.0cm]{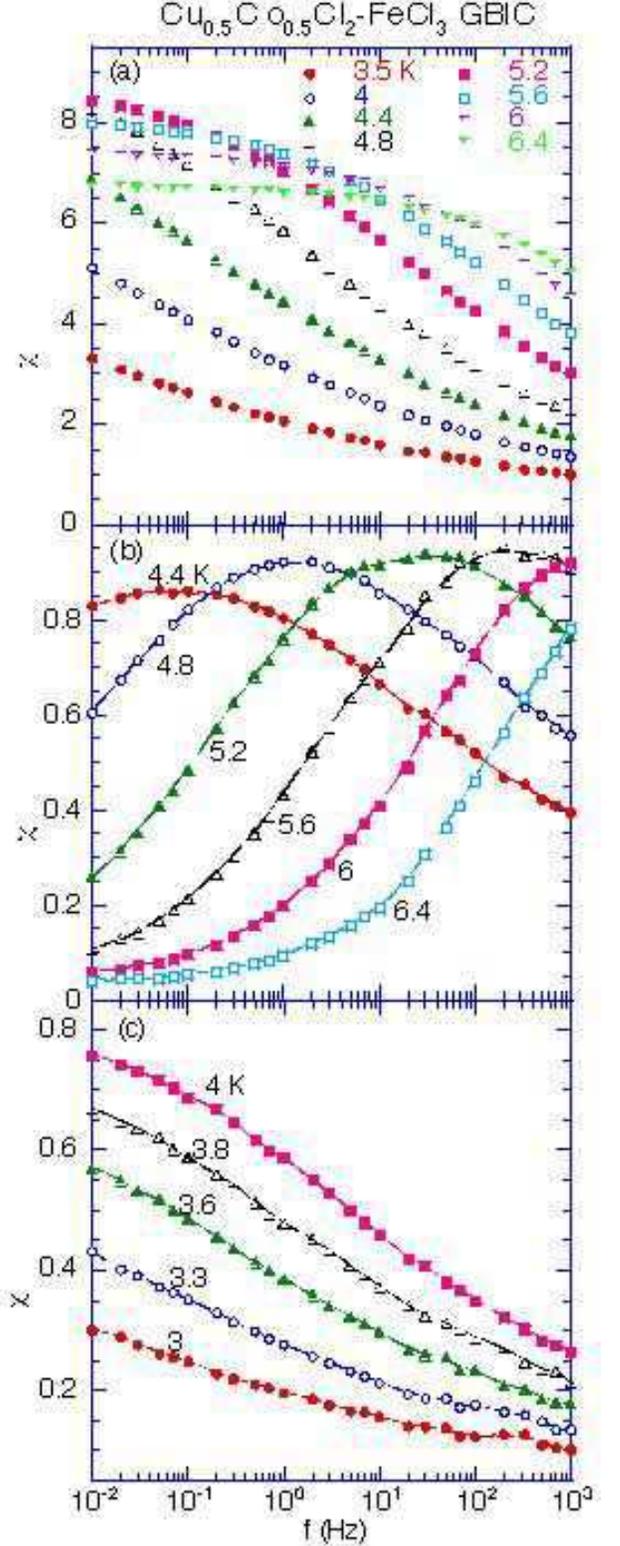}
\caption{\label{fig:six}(a) $f$ dependence of $\chi^{\prime}$ at various
$T$.  $H$ = 0 Oe.  (b) and (c) $f$ dependence of $\chi^{\prime \prime}$
below $T_{g}$ and above $T_{g}$.  $T_{g}$ = 3.92 K. $H$ = 0 Oe.  The 
solid lines are guides to the eyes.}
\end{figure}

Figures \ref{fig:six}(a) and (b) show the $f$ dependence of
$\chi^{\prime}(T,\omega)$ and $\chi^{\prime \prime}(T,\omega)$ at various
$T$ in the vicinity of $T_{g}$ (= 3.92 K), respectively.  The absorption
$\chi^{\prime \prime}(T,\omega)$ curves exhibit different characteristics
depending on $T$.  Above $T_{g}$, $\chi^{\prime \prime}(T,\omega)$ shows a
peak at a characteristic frequency, shifting to the high $f$-side as $T$
increases.  Below $T_{g}$, $\chi^{\prime \prime}(T,\omega)$ shows no peak
for $f \geq$ 0.01 Hz.  It decreases with increasing $f$, following a power
law $\omega^{-\alpha^{\prime \prime}}$.  The exponent $\alpha^{\prime
\prime}$ is almost independent of $T$: $\alpha^{\prime \prime} = 0.096 \pm
0.003$.  According to the fluctuation-dissipation theorem, the magnetic
fluctuation spectrum $S(\omega)$ is related to $\chi^{\prime
\prime}(T,\omega)$ by $S(T,\omega) = 2k_{B}T\chi^{\prime \prime}(T,
\omega)/\omega$.\cite{Svedlindh1989} Then $S(T,\omega)$ is proportional to
$\omega^{-(1+\alpha^{\prime \prime})}$, 
which is similar to 1/$\omega$ character of typical
SG. In contrast, $\chi^{\prime}(T,\omega)$ decreases with increasing $f$ at
any $T$ near $T_{g}$: $\chi^{\prime}$ exhibits a power law
$\omega^{-\alpha^{\prime}}$.  The exponent $\alpha^{\prime}$ is weakly
dependent on $T$: $\alpha^{\prime} = 0.088 \pm 0.001$ at $T$ = 3 K and
$\alpha^{\prime}= 0.111 \pm 0.001$ at $T$ = 3.8 K. The value of
$\alpha^{\prime}$ agrees well with that of $\alpha^{\prime \prime}$.  In
fact, $\chi^{\prime \prime}$ is related to $\chi^{\prime}$ through a so
called ``$\pi/2$ rule'': $\chi^{\prime \prime} =
-(\pi/2)$d$\chi^{\prime}$/d ln$\omega$ (Kramers-Kronig relation), leading
to the relation $\alpha^{\prime} = \alpha^{\prime \prime}$.
Here we note that the observed $f$ dependence of $\chi^{\prime \prime}$ in
Cu$_{0.5}$Co$_{0.5}$Cl$_{2}$-FeCl$_{3}$ GBIC is different from that in
conventional spin glasses such as
Fe$_{0.5}$Mn$_{0.5}$TiO$_{3}$\cite{Gunnarsson1988} and
Eu$_{0.4}$Sr$_{0.6}$S.\cite{Paulsen1987} In these conventional spin
glasses, $\chi^{\prime \prime}$ increases with increasing $f$ both above
$T_{g}$ and below $T_{g}$.

\begin{figure}
\includegraphics[width=7.5cm]{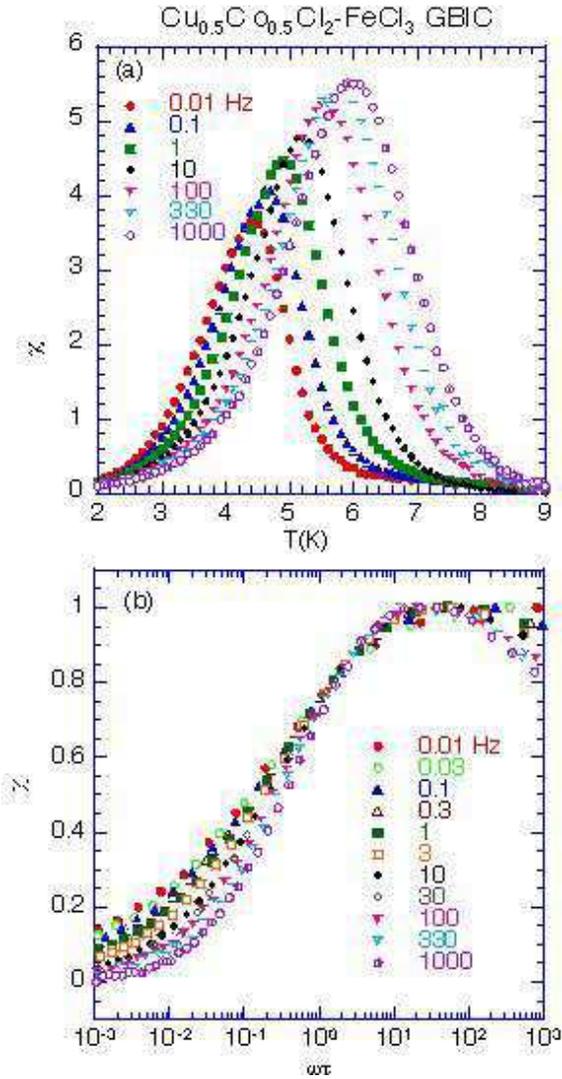}
\caption{\label{fig:seven}(a) $T$ dependence of $T\chi^{\prime \prime}$ at
various $f$.  $H$ = 0.  (b) Scaling plot of $(T\chi^{\prime \prime})_{n}$
(= $(T\chi^{\prime \prime})/(T\chi^{\prime \prime})_{max}$ as a function of
$\omega \tau$.  $(T\chi^{\prime \prime})_{max}$ is the maximum value of
$T\chi^{\prime \prime}$.  The relaxation time $\tau$ is described by Eq. 
(\ref{eq:one}) with $x$ = 10.3, $T_{g}$ = 3.92 K, and $\tau^{*} = 5.27
\times 10^{-6}$ sec.  }
\end{figure}

We consider the validity of a dynamic scaling law for $T\chi^{\prime
\prime}$, which is predicted to be described by\cite{Geschwind1990}
\begin{equation} 
T\chi^{\prime \prime} = \omega^{y}\Omega(\omega \tau),
\label{eq:three} 
\end{equation} 
where $\Omega(\omega \tau)$ is a scaling function of $\omega \tau$ and is
assume to take a maximum at $\omega \tau$ = constant.  The value $y (=
\beta/x)$ is a critical exponent, where $\beta$ is a critical exponent of
the order parameter ($q$).  Figure \ref{fig:seven}(a) shows the $T$
dependence of $T\chi^{\prime \prime}$ at various $f$.  The curve of
$T\chi^{\prime \prime}$ vs $T$ exhibits a peak, which shifts to the
high-$T$ side as $f$ increases.  The peak height of $T\chi^{\prime \prime}$
defined by ($T\chi^{\prime \prime})_{max}$ increases with increasing $f$. 
The least squares fit of the data for the peak height of $T\chi^{\prime
\prime}$ vs $\omega$ (for 0.01 $\leq f \leq$ 1000 Hz) to the form of
($\approx \omega^{y}$) yields $y = 0.035 \pm 0.001$.  Then the value of
$\beta$ (= $xy$) is estimated as $\beta = 0.36 \pm 0.03$, where $x = 10.3
\pm 0.7$.  This value of $\beta$ is smaller than that ($\beta$ = 0.54) for
Fe$_{0.5}$Mn$_{0.5}$TiO$_{3}$.\cite{Gunnarsson1991} Figure
\ref{fig:seven}(b) shows the scaling plot of ($T\chi^{\prime \prime})_{n} =
(T\chi^{\prime \prime})/(T\chi^{\prime \prime})_{max}$ as a function of
$\omega \tau$, where $\tau$ is given by Eq. (\ref{eq:one}) with $x$ =
10.3, $T_{g}$ = 3.92 K, $\tau^{*} = 5.27 \times 10^{-6}$ sec.  In this plot
the data $\chi^{\prime \prime}(T,H=0, \omega)$ for 4 $\leq T \leq$ 10 K and
0.01 Hz $\leq f \leq$ 1 kHz are used.  It seems that the data points fit
well with a scaling function $\Omega(\omega \tau)$, indicating the validity
of the dynamic scaling law given by Eq. (\ref{eq:three}).

\subsection{\label{resultD}Possibility of AT line determined from 
d$\delta$/d$T$ vs $T$}

\begin{figure}
\includegraphics[width=8.0cm]{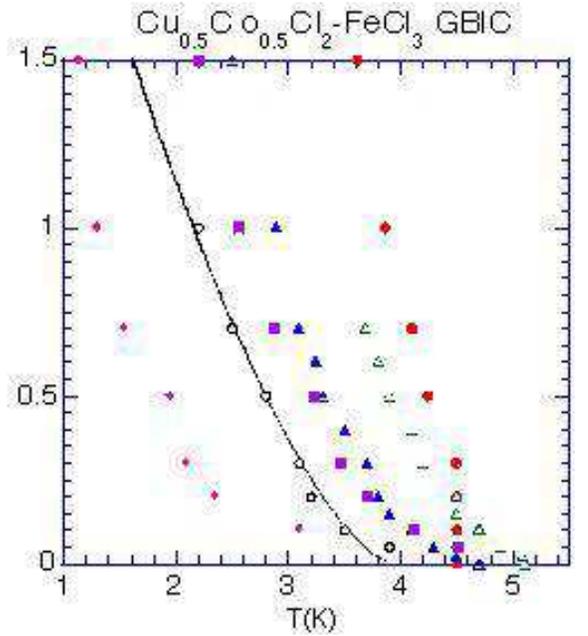}
\caption{\label{fig:eight}$H$-$T$ phase diagram, where the peak
temperatures of $\chi_{ZFC}$ vs $T$ ({\Large $\bullet$}), d$\chi^{\prime
\prime}(T,H,f = 0.1 Hz)$/d$T$ vs $T$ ($\triangle$), $\chi^{\prime \prime}
(T,H,f=0.1$ Hz) vs $T$ ($\blacktriangle$), and the local minimum 
temperature of d$\delta$/d$T$ vs $T$
({\Large $\circ$}).  For comparison, $T_{f}(H,\omega)$ with $f$ = 0.01 Hz
($\blacksquare$) (see also Fig.~\ref{fig:twelve}(a)) and $T_{f}(H)$
($\blacklozenge$) (see Eq. (\ref{eq:eight}) and Fig.~\ref{fig:fourteen}
for definition) are shown as a function of $H$. The solid lines are 
the fits of the data (see text).}
\end{figure}

\begin{figure}
\includegraphics[width=8.0cm]{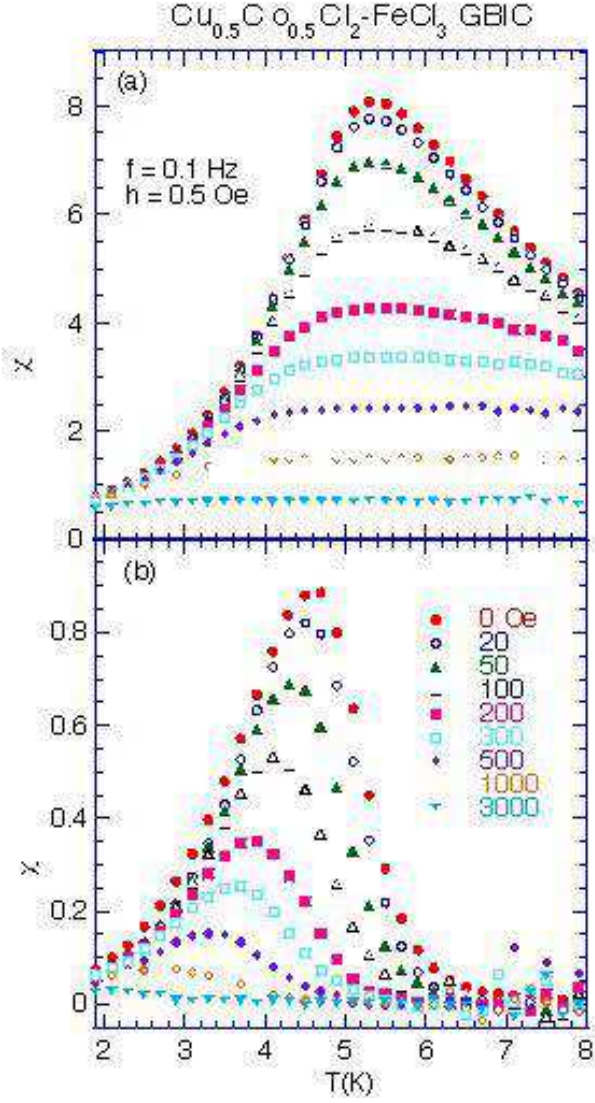}
\caption{\label{fig:nine}$T$ dependence of (a) $\chi^{\prime}$ and (b)
$\chi^{\prime \prime}$ at various $H$.  $f$ = 0.1 Hz and $h$ = 0.5 Oe.}
\end{figure}

According to the AT theory,\cite{Almeida1978} it is predicted that an ideal
SG system in the presence of $H$ undergoes a SG transition at the spin
freezing temperature $T_{f}(H)$ in the $(H,T)$ plane.  The deviation of
$\chi_{ZFC}$ from $\chi_{FC}$ starts to occur for $T < T_{f}(H)$. 
Experimentally it is a little difficult to determine exactly the line
$T_{f}(H)$ at which $\delta = \chi_{FC} - \chi_{ZFC}$ = 0 because of
possible distribution of $T_{f}(H)$ arising from the islandic nature of the
system.  Instead, we use another definition for the line $T_{f}(H)$ at
which d$\delta$/d$T$ vs $T$ exhibits a local minimum at each $H$ (see
Fig.~\ref{fig:two}(b)).  In Fig.~\ref{fig:eight} we show the line
$T_{f}(H)$ thus obtained (the local minimum temperatures of d$\delta$/d$T$
vs $T$) in the $(H,T)$ plane.  For comparison, we also plot the peak
temperatures of $\chi_{ZFC}(T, H)$ vs $T$, $\chi^{\prime
\prime}(T,H,\omega)$ vs $T$ at $f$ = 0.1 Hz, and the local minimum 
temperature of d$\chi^{\prime
\prime}(T,H,\omega)$/d$T$ vs $T$ with $f$ = 0.1 Hz as a function of $H$ in
the $(H,T)$ plane.  Note that the data of $\chi^{\prime
\prime}(T,H,\omega)$ vs $T$ at $f$ = 0.1 Hz will be given later (see
Fig.~\ref{fig:nine}).  These lines are away from the line $T_{f}(H)$.  The
least squares fit of the data of the line $T_{f}(H)$ for 100 Oe $\leq H
\leq$ 1 kOe to
\begin{equation} 
H = H_{0}[1-T_{f}(H)/T_{g}]^{p}, 
\label{eq:four} 
\end{equation} 
yields parameters $p = 1.52 \pm 0.10$ and $H_{0} = (3.4 \pm 0.4)$ kOe,
where $T_{g} = 3.92 \pm 0.11$ K. In the mean field
picture,\cite{Marinari1998,Almeida1978} this line corresponds to the AT
line.  In fact, the value of exponent $p$ is close to that ($p$ = 1.50) for
the AT line.

It has been believed that the mean field picture is valid for
Fe$_{0.5}$Mn$_{0.5}$TiO$_{3}$: $p$ = 1.49.\cite{Katori1994} However,
Mattsson et al.\cite{Mattsson1995} have claimed that the SG transition is
destroyed by the application of a finite magnetic field $H$ in
Fe$_{0.5}$Mn$_{0.5}$TiO$_{3}$.  Their result supports the prediction from
the scaling picture that the long range SG order at equilibrium only occurs
for $H$ = 0 and $T < T_{g}$.  Further discussion for the AT line will be
done in Secs.  \ref{resultE} and \ref{resultF}.

\subsection{\label{resultE}Effect of $H$ on the SG transition}

\begin{figure}
\includegraphics[width=8.5cm]{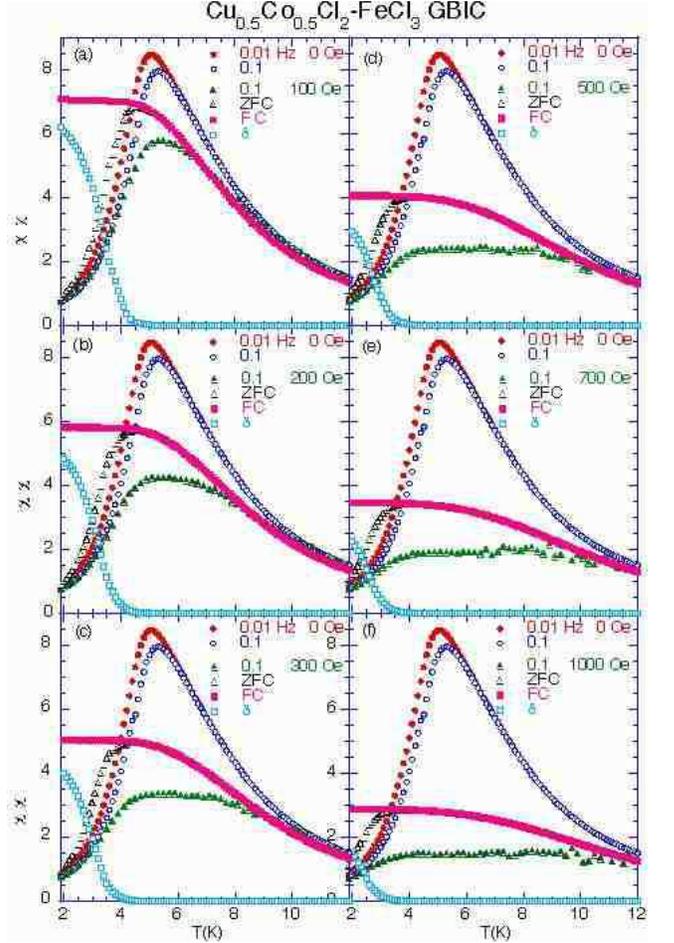}
\caption{\label{fig:ten}(a) - (f) $T$ dependence of $\chi^{\prime}$ at $f$
= 0.01 and 0.1 Hz ($H$ = 0), $\chi^{\prime}$ at $f$ = 0.1 Hz at $H$,
$\chi_{ZFC}$, $\chi_{FC}$ and $\delta$ (= $\chi_{FC} - \chi_{ZFC}$) at the
same $H$.  $H$ = 100, 200, 300, 500, 700, and 1000 Oe.}
\end{figure}

\begin{figure}
\includegraphics[width=8.0cm]{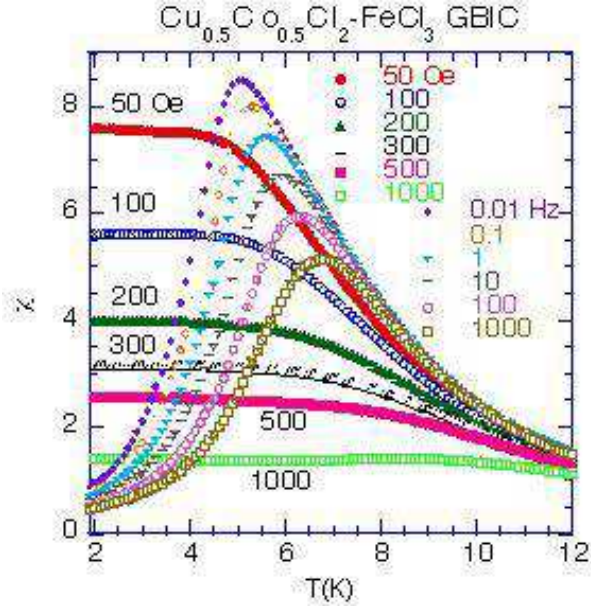}
\caption{\label{fig:eleven}$T$ dependence of $\chi^{\prime}(T,H=0,\omega)$
and d$M_{FC}(T,H)/$d$H$.  The spin freezing temperature $T_{f}(H,\omega)$ is
defined by a temperature at which $\chi^{\prime}(T,H=0,\omega)$ is equal to
d$M_{FC}(T,H)/$d$H$.}
\end{figure}

\begin{figure}
\includegraphics[width=8.0cm]{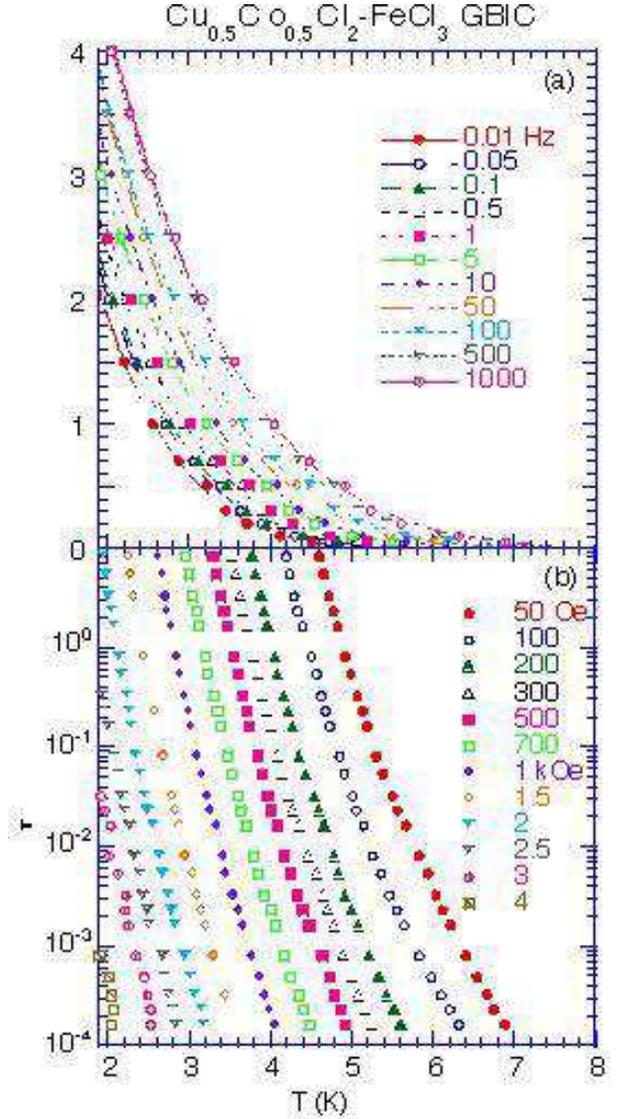}
\caption{\label{fig:twelve}(a) $T_{f}(H,\omega)$ contours at various $H$
between $f$ = 0.01 Hz and 1 kHz, which is determined from
Fig.~\ref{fig:eleven}.  (b) $\tau$ vs $T$ at various $H$, which is derived
from (a). The lines are guides to the eyes.}
\end{figure}

The transition from the paramagnetic phase to the SG phase occurs in our
system at $H$ = 0.  What happens to this transition in the presence of $H$? 
Does the long-range SG phase survive as suggested by the molecular field
picture or is destroyed as suggested by the scaling picture?  According to
the method developed by Mattsson et al.,\cite{Mattsson1995} here we have
determined the dynamic freezing temperature $T_{f}(H,\omega)$ in a wide
frequency and field interval.  Figure \ref{fig:nine} shows the $T$
dependence of $\chi^{\prime}(T,H,\omega)$ and $\chi^{\prime
\prime}(T,H,\omega)$ at various $H$ for $f = \omega/2\pi$ = 0.1 Hz, where
$h$ = 0.5 Oe and $H$ (0 $< H \leq$ 3 kOe) is applied along the $c$ plane
perpendicular to the $c$ axis.  In Figs.~\ref{fig:ten}(a) - (f), we show
the $T$ dependence of $\chi^{\prime}(T,H=0,\omega)$ at $f$ = 0.01 and 0.1 Hz,
$\chi^{\prime}(T,H,\omega)$ at $f$ = 0.1 Hz, $\chi_{FC}(T,H)$,
$\chi_{ZFC}(T,H)$, and $\delta = \chi_{FC}(T,H) - \chi_{ZFC}(T,H)$ at
various $H$ = 100 - 1000 Oe.  
The susceptibility $\chi_{FC}(T,H)$ (= $M_{FC}/H$) corresponds to the
equilibrium susceptibility.  Since $M_{FC}(T,H)$ is a non-linear function
of $H$, $\chi_{FC}$ (= $M_{FC}/H$) is not equal to the differential FC
susceptibility d$M_{FC}/$d$H$.  The numerical calculations indicates that
d$M_{FC}/$d$H$ is smaller than $\chi_{FC}$ at the same $H$ and
$T$.  In Fig.~\ref{fig:eleven} we show the $T$ dependence of d$M_{FC}(T,H)/$d$H$ at
various $H$ and $\chi^{\prime}(T,H=0,\omega)$ at various $f$.  The spin
freezing temperature $T_{f}(H,\omega)$ is defined as a temperature at which
d$M_{FC}(T,H)/$d$H$ coincides with $\chi^{\prime}(T,H=0,\omega)$. 
As shown in Fig.~\ref{fig:ten}, $\chi^{\prime}(T,H,\omega)$ at $f=0$.  1 Hz deviates
from $\chi^{\prime}(T,H=0,\omega)$ at the same $f$ above $T_{f}(H,\omega)$.  This
is the same method that has been used by Mattsson et al.\cite{Mattsson1995}
to obtain $T_{f}(H,\omega)$ for Fe$_{0.5}$Mn$_{0.5}$TiO$_{3}$.  The
advantage of this method is that we do not have to measure the $T$
dependence of $\chi^{\prime}$ and $\chi^{\prime \prime}$ at each $H$ such
as the data shown in Fig.~\ref{fig:nine}.  In Fig.~\ref{fig:twelve}(a) we
show a plot of the line $T_{f}(H,\omega)$ in the $(H,T)$ plane, where $f$
(0.01 $\leq f \leq$ 1000 Hz) is varied as a parameter.  
For comparison, the line $T_{f}(H,\omega)$ at $f$ = 0.01 Hz is also plotted
in the $H$-$T$ diagram of Fig.~\ref{fig:eight}.  We find that this line is
close to the line $T_{f}(H)$ determined from the local minimum temperature
of d$\delta/$d$T$ vs $T$.  This result suggests that the method to
determine $T_{f}(H,\omega)$ is appropriate.  Note that similar
data of $T_{f}(H,\omega)$ has been reported by Mattsson et al.  for
Fe$_{0.5}$Mn$_{0.5}$TiO$_{3}$.\cite{Mattsson1995} Figure
\ref{fig:twelve}(b) shows the $T$ dependence of $\tau(T,H)$ which is
derived from Fig.~\ref{fig:twelve}(a); $\tau(T,H) = 1/\omega = 1/(2\pi f)$
at $T= T_{f}(H,\omega)$.  

Assuming a SG phase transition at $T_{g}$, we assume the following dynamic
scaling relation for the relaxation time $\tau_{\pm}(T,H)$ in the presence
of $H$,\cite{Fischer1991}
\begin{equation} 
\tau_{\pm}(T,H)=\mid \epsilon \mid^{-x} F_{\pm}(X) 
= H^{-2x/(\beta + \gamma)} G_{\pm}(X)
\label{eq:five} 
\end{equation} 
with $X = \mid \epsilon \mid H^{-2/(\beta + \gamma)}$,
where ($+$) denotes $T > T_{g}$ and ($-$) denotes $T < T_{g}$, $\epsilon$
is the reduced temperature defined by $\epsilon = T/T_{g} - 1$, $\tau_{c}
\approx \tau_{0} \mid \epsilon \mid^{-x}$ is the relaxation time
for $T > T_{g}$ at $H$ = 0, $\gamma$ is the exponent of nonlinear dynamic
susceptibility $\chi_{3}$, $F_{+}$ and $G_{+}$ are the scaling functions
for $T > T_{g}$ and $F_{-}$ and $G_{-}$ are ones for $T < T_{g}$.  The
scaling relation Eq. (\ref{eq:five}) suggests that $\tau$ diverges like
$\tau(T=T_{g},H) \approx H^{-2x/(\beta+\gamma)}$ as $H \rightarrow$ 0.  The
characteristic field denoted by $H^{*} \approx \mid \epsilon
\mid^{(\beta+\gamma)/2}$ is a crossover line between a weak-field and a
strong-field region for $T > T_{g}$.

In the scaling picture, the asymptotic form for $\tau_{-}$ below $T_{g}$ is
obtained as follows.  In Eq. (\ref{eq:two}) we assume that $l$ diverges
like $l \approx H^{-2/(d-2\theta)}$ as $H$ reduces to
zero.\cite{Fisher1988b} Then $\tau_{-}$ can be rewritten as
\begin{equation} 
\tau_{-}(T,H) \approx H^{-2z/(d-2\theta)} \exp[Y_{0}\mid \epsilon
\mid^{\psi \nu} H^{-2\psi/(d-2\theta)}/k_{B}T_{g}],
\label{eq:six} 
\end{equation} 
just below $T_{g}$, where $\theta$ is the energy exponent.  The equivalence
of Eqs. (\ref{eq:five}) and (\ref{eq:six}) leads to the relation $\nu =
(\beta+\gamma)/(d-2\theta)$.  Since $\beta + \gamma = \nu (d - 2\theta)=
2-\alpha-2\nu \theta=2\beta + \gamma- 2\nu \theta$ , we obtain a scaling
relation ($\beta = 2\nu \theta$).  Then $F_{-}(X)$ in Eq. (\ref{eq:five})
has the asymptotic form of $F_{-}(X) \approx \exp(X^{\psi \nu})$, 
since $Y_{0}/(k_{B}T_{g}) \approx 1$.

Our result of $\tau(T,H)$ shown in Fig.~\ref{fig:twelve}(b) is analyzed
using the dynamic scaling relation.  The relaxation time $\tau(T=T_{g},H)$
is predicted to be proportional to $H^{-2x/(\beta+\gamma)}$ as $H
\rightarrow 0$.  In fact, the least squares fit of the data of $\tau$ vs
$H$ at $T_{g}$ = 3.92 K for 0.1 $\leq H \leq$ 1 kOe yields the parameters
$2x/(\beta +\gamma) = 5.30 \pm 0.13$.  Since $x = 10.3 \pm 0.7$ and $\beta
= 0.36 \pm 0.03$, the exponent $\gamma$ is estimated as $\gamma = 3.5 \pm
0.4$.  Using the scaling relations ($\alpha + 2\beta + \gamma$ = 2 and 2 -
$\alpha = d\nu)$, the exponents $\alpha$ and $\nu$ are calculated as
$\alpha = -2.2 \pm 0.5$ and $\nu = 1.4 \pm 0.2$, where $d$ = 3.  The
exponent $z$ is given by $z = 6.6 \pm 1.2$.  These results agree with those
predicted from the MC simulation by Ogielski for the 3D $\pm J$ Ising model
($\alpha = -1.9 \pm 0.3$, $\beta$ = 0.5, $\gamma = 2.9 \pm 0.3$, $\nu = 1.3
\pm 0.1$, $z = 6.0 \pm 0.5)$.\cite{Ogielski1985} The exponent $\theta$ is
estimated as $\theta = 0.13 \pm 0.02$ using the scaling relation ($\theta =
\beta/2\nu$).  The validity of this $\theta$ will be discussed in
Sec.~\ref{dis}.

\begin{figure}
\includegraphics[width=8.0cm]{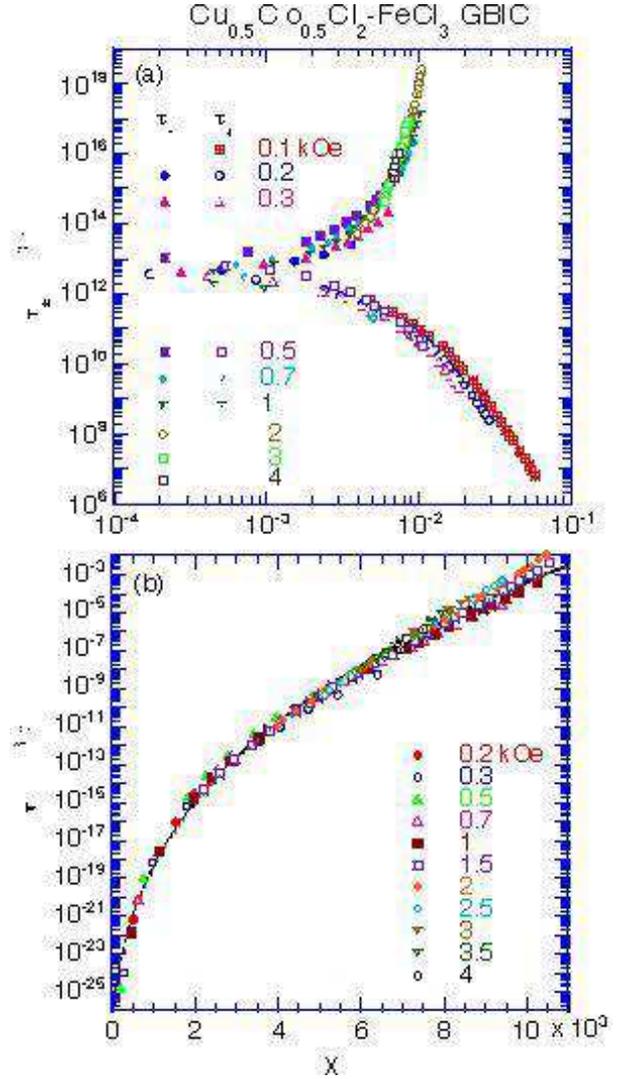}
\caption{\label{fig:thirteen}(a) Scaling plot of $\tau_{\pm}
H^{2x/(\beta+\gamma)}$ vs $X$ at various $H$, where $X = \mid \epsilon
\mid H^{-2/(\beta+\gamma)}$.  The values of exponents $\beta$,
$\gamma$, and $x$ are given in the text.  (b) Scaling plot of $X^{x}$
$\tau_{-} H^{2x/(\beta+\gamma)}$ vs $X$ at various $H$.  The solid line is
a fitting curve (see the text for detail).}
\end{figure}

Figure \ref{fig:thirteen}(a) shows a scaling plot of
$\tau_{\pm}(T,H)H^{2x/(\beta+\gamma)}$ a function of $X$,
\begin{equation} 
\tau_{\pm}(T,H)H^{2x/(\beta+\gamma)} = G_{\pm}(X),
\label{eq:seven} 
\end{equation} 
where $x$ = 10.3, $T_{g}$ = 3.92 K, $\beta = 0.36$, and $\gamma =
3.5$.  We find that almost all the data fall well on two scaling
functions: $G_{+}$ for $T > T_{g}$ and $G_{-}$ for $T < T_{g}$.  The
scaling function $F_{\pm}$ is related to the scaling function $G_{\pm}$ by
$F_{\pm}(X) = X^{x}G_{\pm}(X)$ with $x$ = 10.3.  In
Fig.~\ref{fig:thirteen}(b) we show the plot of $F_{\pm}(X)$ as a function
of $X$.  The form of $F_{-}(X)$ is given as follows.  $\ln [F_{-}(X)] =
c_{1}X^{\psi \nu} + c_{2}$ with $c_{1} = 300 \pm 18$, $c_{2} = -72 \pm 2$,
and $\psi \nu = 0.33 \pm 0.01$.  This fitting curve is determined from the
data of $\tau_{-}$ at $H$ = 3 kOe.  Since $\nu = 1.4 \pm 0.2$, $\psi$ is
estimated as 0.24 $\pm$ 0.02.

\subsection{\label{resultF}Possibility of SG phase transition at finite $H$} 

\begin{figure}
\includegraphics[width=8.0cm]{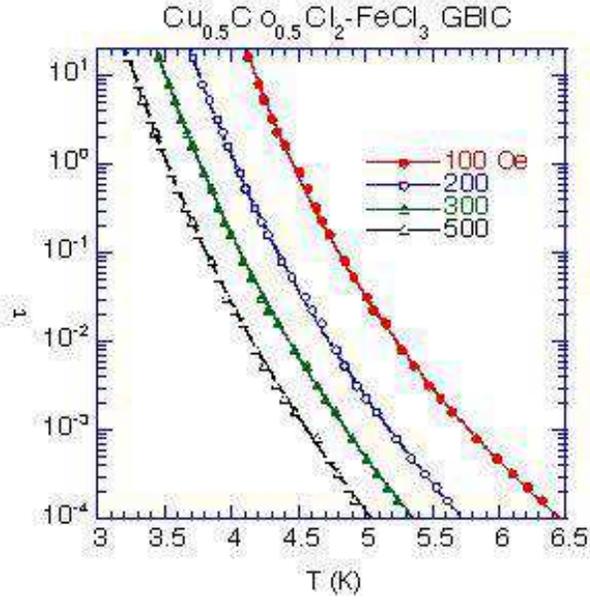}
\caption{\label{fig:fourteen}Plot of $\tau$ vs $T$ obtained from the data
of $T_{f}(H,\omega)$ with $H$ = 100, 200, 300, and 500 Oe.  The relaxation
time $\tau$ is defined as $\omega \tau$ = 1 at $T = T_{f}(H,\omega)$.}
\end{figure}

We consider the possibility that $\tau(T,H)$ may be described by a critical
slowing down given by
\begin{equation} 
\tau(T,H) = \tau^{*}[T/T_{f}(H)-1]^{-x}
\label{eq:eight} 
\end{equation} 
for $T > T_{f}(H)$.  This is based on the assumption that the SG transition
is not destroyed by the application of $H$ (the molecular field picture). 
The SG phase exists below $T_{f}(H)$ in thermal equilibrium.  The least
squares fits of the data of $\tau(T,H)$ vs $T$ shown in
Fig.~\ref{fig:fourteen} to the above power law form yield $x$ = 10.2 $\pm$
0.3, $T_{f}(H)$ = 3.10 $\pm$ 0.05 K, $\tau^{*}$ = 2.3 $\times$ 10$^{-4}$
sec at $H$ = 100 Oe, $x$ = 13.2 $\pm$ 0.4, $T_{f}(H)$ = 2.35 $\pm$ 0.07 K,
$\tau^{*}$ = 1.2 $\times$ 10$^{-2}$ sec at $H$ = 200 Oe, $x$ = 13.8 $\pm$
0.5, $T_{f}(H)$ = 2.08 $\pm$ 0.08 K, $\tau^{*}$ = 5.2 $\times$ 10$^{-2}$
sec at $H$ = 300 Oe, and $x$ = 13.6 $\pm$ 0.6, $T_{f}(H)$ = 1.94 $\pm$ 0.08
K, $\tau^{*}$ = 6.0 $\times$ 10$^{-2}$ sec at $H$ = 500 Oe, The values of
$T_{f}(H)$ thus obtained are plotted as a function of $H$ in
Fig.~\ref{fig:eight}.  This line does not coincide with the line $T_{f}(H)$
determined form the local minimum temperature of d$\delta$/d$T$ vs $T$. 
These two lines start to deviate at $H \approx$ 0 Oe.  The exponent $x$ is
strongly dependent on $H$.  The value of $x$ at $H$ = 100 Oe is almost
equal to $x$ at $H$ = 0, but it drastically increases with increasing $H$
for $H \geq$ 200 Oe.  These results suggest that the SG transition at $H$ =
0 is destroyed by the application of $H$ (at least above 100 Oe).  This
result is consistent with the following theoretical prediction proposed by
Lamarcq et al.\cite{Lamarcq2002} For $H<H_{0}$ (some characteristic field)
the system behaves qualitatively just as in the case $H$ = 0, while
significant changes arise for $H>H_{0}$).  Our result is slightly different
from the conclusion derived by Mattsson et al.\cite{Mattsson1995} that the
SG phase is destroyed at finite $H$ for Fe$_{0.5}$Mn$_{0.5}$TiO$_{3}$. 
Note that there are some recent theories\cite{Houdayer1999} supporting the
result of Mattsson et al.

\subsection{\label{resultG}Aging at constant temperature}

\begin{figure}
\includegraphics[width=8.0cm]{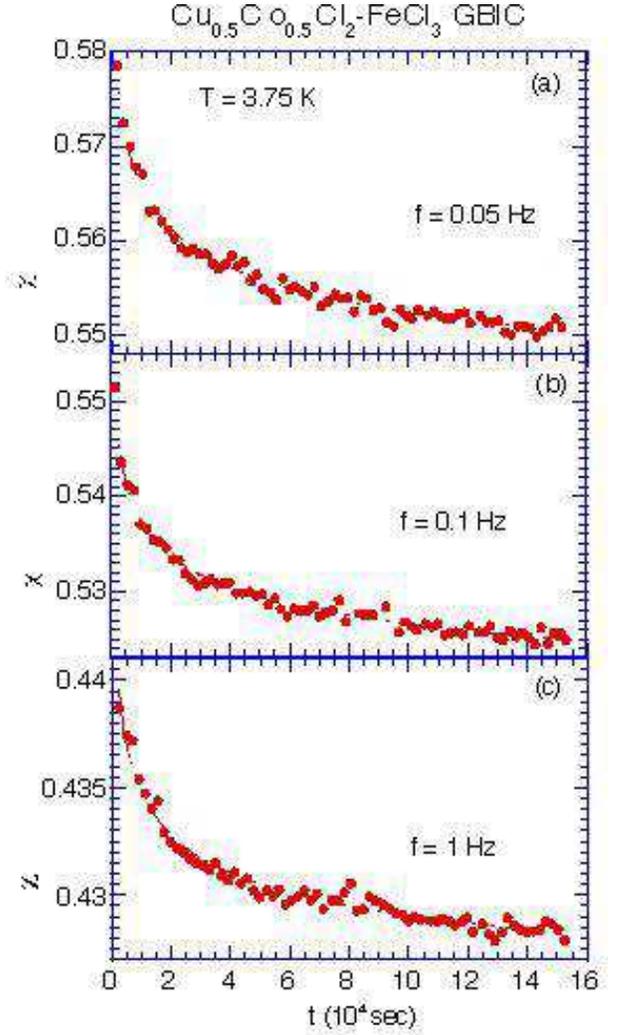}
\caption{\label{fig:fifteen}$t$ dependence of $\chi^{\prime \prime}$ at $T$
= 3.75 K for $f$ = 0.05, 0.1, and 1 Hz, where $t$ is the time taken after
the sample was quenched from 50 to 3.75 K. The solid lines are the 
fits of the data to the power law form.}
\end{figure}

\begin{figure}
\includegraphics[width=8.0cm]{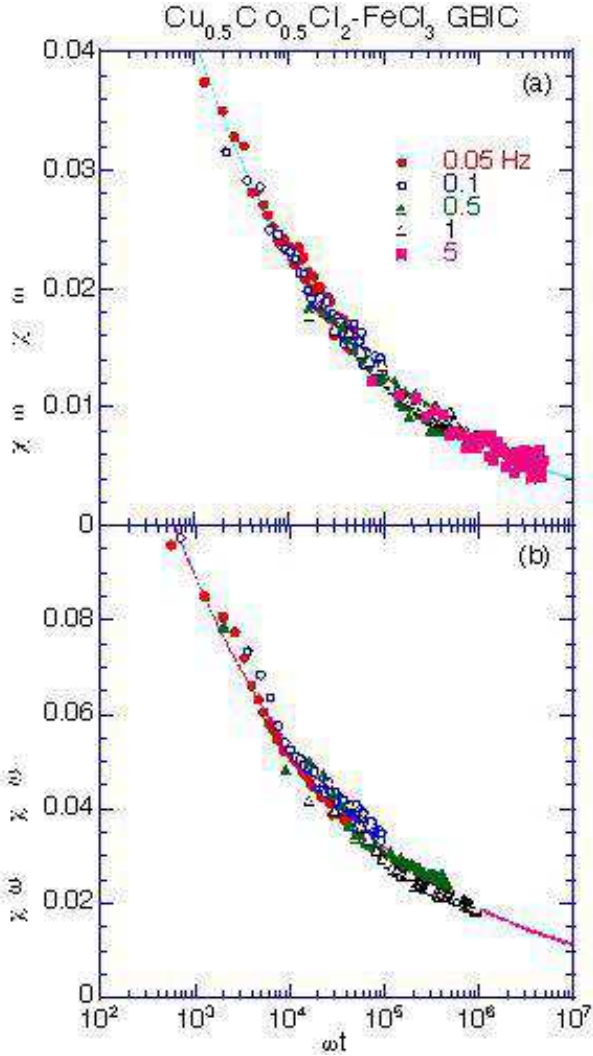}
\caption{\label{fig:sixteen}Scaling plot of (a) $\Delta \chi^{\prime
\prime}(\omega,t)$ [= $\chi^{\prime \prime}(\omega,t) - \langle
\chi^{\prime \prime}_{0}(\omega)\rangle$] and (b) $\Delta
\chi^{\prime}(\omega,t)$ [= $\chi^{\prime}(\omega,t) - \langle
\chi^{\prime}_{0}(\omega)\rangle$] as a function of $\omega t$ for the data
at $f$ = 0.05, 0.1, 0.5, 1, and 5 Hz.  The definition of stationary
susceptibility $\langle \chi^{\prime \prime}_{0}(\omega)\rangle$ and
$\langle \chi^{\prime}_{0}(\omega)\rangle$ is given in the text.  The solid
lines are the fits of the data to the power law form.  Each curve is
vertically shifted by $\langle \chi^{\prime \prime}_{0}(\omega)\rangle$ or
$\langle \chi^{\prime}_{0}(\omega)\rangle$ (see text for definition).}
\end{figure}

\begin{table}
\caption{\label{table:one}Fit parameters for the decay of $\chi^{\prime
\prime}$ vs $t$.  $T$ = 3.75 K to Eq. (\ref{eq:nine}).  The definition of
the parameters is given in the text.}
\begin{ruledtabular}
\begin{tabular}{lllll}
    $f$ & $b^{\prime \prime}$ & $\chi^{\prime 
    \prime}_{0}(\omega)$ & $\langle \chi^{\prime 
    \prime}_{0}(\omega)\rangle$ & $A^{\prime \prime}(\omega)$\\
    (Hz) &  & (emu/av mol) &  &  \\
0.05 & 0.159 $\pm$ 0.022 & 0.522 $\pm$ 0.005 & 0.535 & 0.188 $\pm$ 0.016\\
0.1 & 0.158 $\pm$ 0.019 & 0.502 $\pm$ 0.004 & 0.512 & 0.152 $\pm$ 0.010\\
0.5 & 0.185 $\pm$ 0.021 & 0.445 $\pm$ 0.002 & 0.449 & 0.111 $\pm$ 0.010\\
1 & 0.193 $\pm$ 0.033 & 0.419 $\pm$ 0.002 & 0.420 & 0.095 $\pm$ 
0.016\\
\end{tabular}
\end{ruledtabular}
\end{table}

We have measured the $t$ dependence of $\chi^{\prime \prime}$ at $T$ =
3.25, 3.50, 3.75, 3.90, and 4.4 K, where $H$ = 0.  The system was quenched
from 10 K to $T$ ($< T_{g}$) at time (age) zero.  Both $\chi^{\prime}$ and
$\chi^{\prime \prime}$ were measured simultaneously as a function of time
$t$ at constant $T$.  Each point consists in the successive measurements of
five frequencies ($f$ = 0.05, 0.1, 0.5, 1, and 5 Hz).  Figure
\ref{fig:fifteen} shows the $t$ dependence of $\chi^{\prime \prime}$ at $T$
= 3.75 K for $f$ = 0.05, 0.1, and 1 Hz.  The absorption $\chi^{\prime
\prime}$ decreases with increasing $T$ and is well described by a power-law
decay
\begin{equation} 
\chi^{\prime \prime}(\omega,t) = \chi^{\prime \prime}_{0}(\omega)+A^{\prime
\prime}(\omega)t^{-b^{\prime \prime}}.
\label{eq:nine} 
\end{equation} 
The least squares fit of the data of $\chi^{\prime \prime}(\omega,t)$ at
$T$ = 3.75 K to Eq. (\ref{eq:nine}) yields fitting parameters listed in
Table \ref{table:one}.  The exponent $b^{\prime \prime}$ is slightly
dependent on $f$.  The $f$ dependence of the amplitude $A^{\prime
\prime}(\omega)$ for 0.05 $\leq f \leq$ 1 Hz is described by $A^{\prime
\prime}(\omega) = A_{0}^{\prime \prime}\omega^{-\mu^{\prime \prime}}$
with $A_{0}^{\prime \prime} = 0.142 \pm 0.002$ and $\mu^{\prime \prime} =
0.225 \pm 0.016$.  The value of $\mu^{\prime \prime}$ is almost the same as
that of $b^{\prime \prime}$, supporting a $\omega t$-scaling relation that
Eq. (\ref{eq:nine}) can be rewritten as
\begin{equation} 
\chi^{\prime \prime}(\omega,t) = 
\chi^{\prime \prime}_{0}(\omega) 
+A_{0}^{\prime \prime}(\omega t)^{-b^{\prime \prime}}.
\label{eq:ten} 
\end{equation} 
In Fig.~\ref{fig:sixteen}(a) we show a scaling plot of $\Delta \chi^{\prime
\prime}(\omega,t)$ at 3.75 K as a function of $\omega t$, where $\Delta
\chi^{\prime \prime}(\omega,t) = \chi^{\prime \prime}(\omega,t) - \langle
\chi^{\prime \prime}_{0}(\omega)\rangle$.  The stationary susceptibility
$\langle \chi^{\prime \prime}_{0}(\omega)\rangle$ listed in Table
\ref{table:one} is slightly different from $\chi^{\prime
\prime}_{0}(\omega)$ and corresponds to the asymptotic $f$-dependent value
so that $\chi^{\prime \prime}(\omega,t)$ tends to zero in the limit of $t
\rightarrow \infty$.  Almost all the data fall well on a single scaling
function given by Eq. (\ref{eq:ten}) with $b^{\prime \prime}= \langle
b^{\prime \prime}\rangle = 0.255 \pm 0.005$ and $A^{\prime \prime}_{0} =
\langle A^{\prime \prime}_{0} \rangle = 
0.239 \pm 0.009$.  The value of $\langle b^{\prime \prime}\rangle$ is a
little larger than that of $b^{\prime \prime}$ determined
using the data at $f$ = 0.05 Hz.  Note that $\langle b^{\prime
\prime}\rangle$ is equal to 0.14 $\pm$ 0.03 at 19 K for
Fe$_{0.5}$Mn$_{0.5}$TiO$_{3}$ ($T_{g}$ = 20.7 K).\cite{Dupuis2001} The
susceptibility $\langle \chi^{\prime \prime}_{0}(\omega)\rangle$ clearly
decreases with increasing $\omega$, which is consistent with the prediction
by Picco et al.\cite{Picco2001} 

In Fig.~\ref{fig:sixteen}(b) we show a scaling plot of $\Delta
\chi^{\prime}(\omega,t)$ at $T$ = 3.75 K, where $\Delta
\chi^{\prime}(\omega,t) = \chi^{\prime}(\omega,t) - \langle
\chi^{\prime}_{0}(\omega)\rangle$ and $\langle
\chi^{\prime}_{0}(\omega)\rangle$ is the stationary part of
$\chi^{\prime}(\omega,t)$ in the limit of $t \rightarrow \infty$.  All the
data lie well with a universal curve described by $\Delta
\chi^{\prime}(\omega,t) = A_{0}^{\prime}(\omega t)^{-b^{\prime}}$ with
$b^{\prime} = \langle b^{\prime}\rangle = 0.226 \pm 0.007$ and
$A_{0}^{\prime} = 0.426 \pm 0.018$.  The exponent $\langle b^{\prime}
\rangle$ is nearly equal to $\langle b^{\prime \prime} \rangle$.  The
susceptibility $\langle \chi^{\prime}_{0}(\omega) \rangle$ decreases with
increasing $f$, which is similar to the case of $\langle \chi^{\prime
\prime}_{0}(\omega)\rangle$.

\begin{figure}
\includegraphics[width=8.0cm]{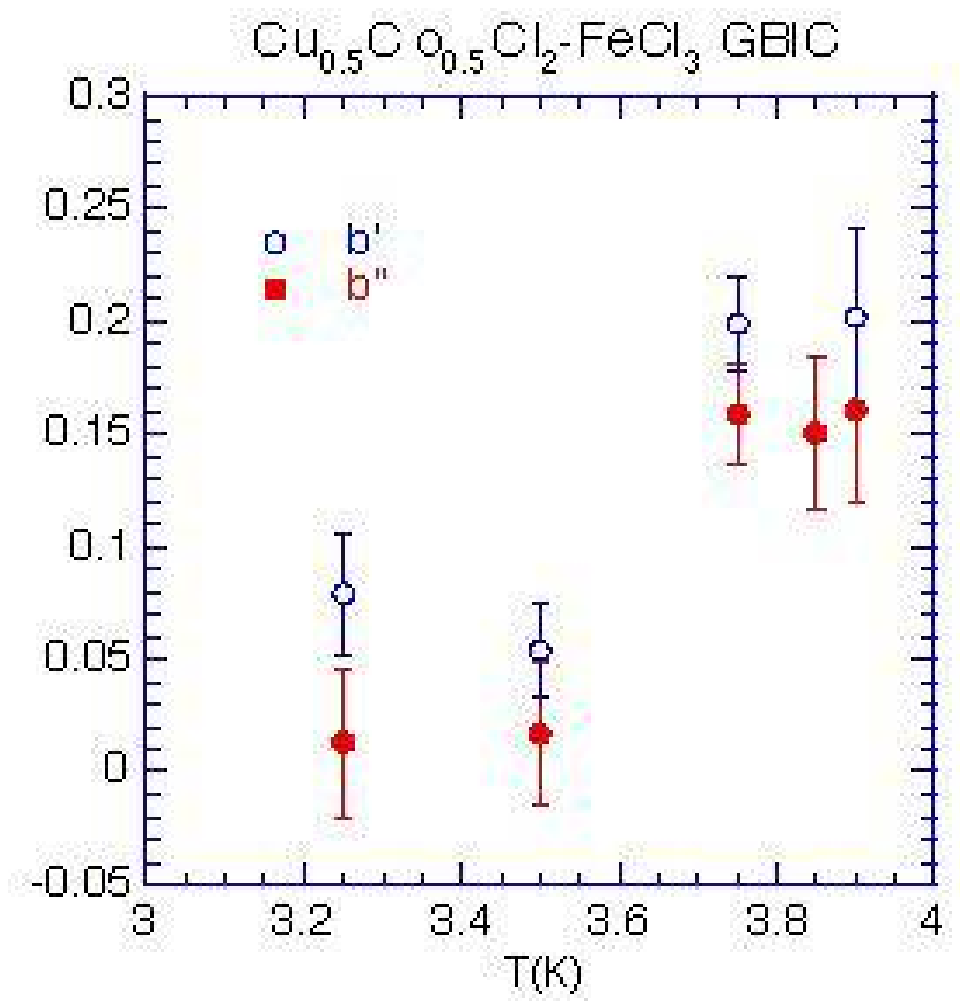}
\caption{\label{fig:seventeen}$T$ dependence of exponents $b^{\prime}$ and 
$b^{\prime \prime}$.}
\end{figure}

Such a power-law decay of $\chi^{\prime \prime}(\omega, t)$ and
$\chi^{\prime}(\omega,t)$ with $t$ is observed only for 3.25 K $\leq T \leq
T_{g}$.  In Fig.~\ref{fig:seventeen} we shows the $T$ dependence of
$b^{\prime}$ and $b^{\prime \prime}$ at $f$ = 0.05 Hz.  Both $b^{\prime}$
and $b^{\prime \prime}$ are on the same order at the same $T$.  There is a
step-like increase in $b^{\prime}$ and $b^{\prime \prime}$ with increasing
\textit{T} between 3.5 and 3.75 K; $b^{\prime \prime} = 0.017 \pm 0.032$ at
3.50 K and $b^{\prime \prime} = 0.16 \pm 0.04$ at 3.9 K just below $T_{g}$. 
Similar behavior on the $T$ dependence of $b^{\prime \prime}$ has been
reported by Colla et al.\cite{Colla2000} for a relaxor ferroelectric
Pb(Mg$_{1/3}$Nb$_{2/3}$)O$_{3}$, which is not a SG system.  We note that as
shown in Fig.~\ref{fig:two}(b), the data of d$\delta$/d$T$ vs $T$ for $H$ =
1 Oe exhibits a local minimum at 3.5 K, close to a temperature at which
$b^{\prime \prime}$ undergoes a drastic increase.

\subsection{\label{resultI}Rejuvenation effect under perturbations}

\begin{figure}
\includegraphics[width=8.0cm]{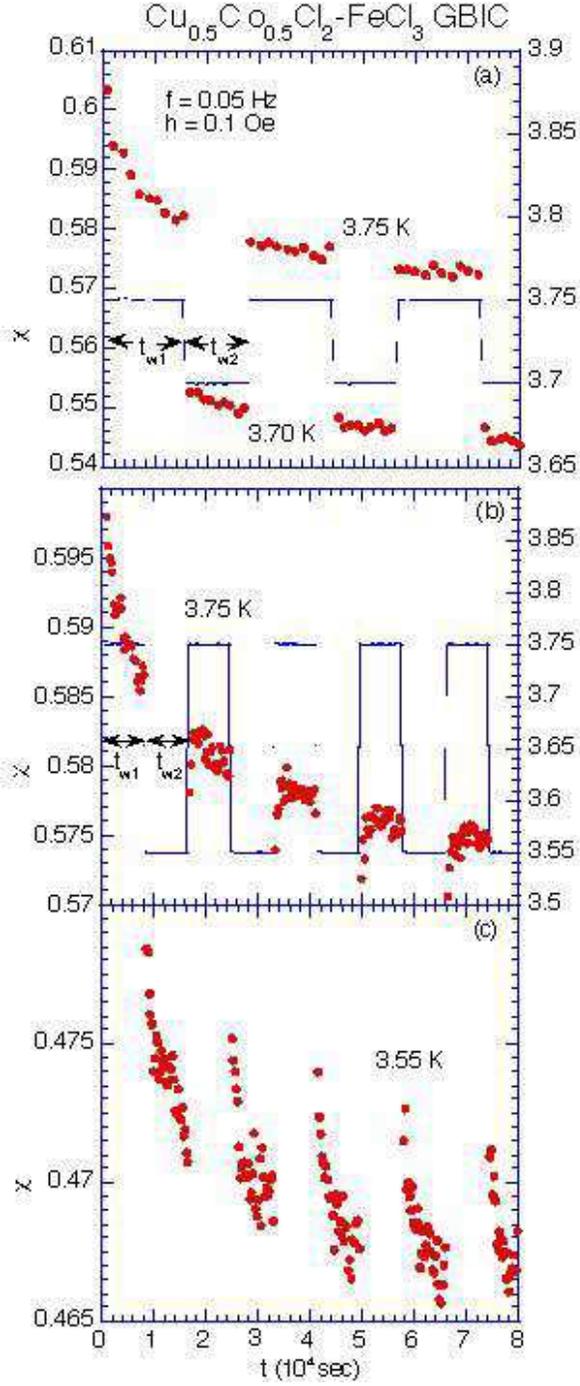}
\caption{\label{fig:eighteen}Relaxation of $\chi^{\prime \prime}(\omega,t)$
during a negative temperature cycle.  $f$ = 0.05 Hz, $T$ = 3.75 K and $T -
\Delta T$. $H$ = 0.  The change of $T$ with $t$ is also shown. (a) 
$\Delta T$ = 0.05 K. (b) and (c) $\Delta T$ = 0.20 K.}
\end{figure}

We also examine the aging effect of $\chi^{\prime \prime}$ under 
perturbations such as the change of $T$ and $H$. Figure \ref{fig:eighteen}
shows the $t$ dependence of $\chi^{\prime \prime}(\omega,t)$ under a
negative temperature-cycle, where $f$ = 0.05 Hz and $h$ = 0.1 Oe.  Here our
system was quenched from 10 to $T$ = 3.75 K at $H$ = 0.  The relaxation
of $\chi^{\prime \prime}(\omega,t)$ was measured as a function of $t$
during a period $t_{w1}$ ($\approx 1.5 \times 10^{4}$ sec).  The
temperature was then changed to $T-\Delta T$ (the first
$T$-shift).  The relaxation of $\chi^{\prime \prime}$ was measured as a
function of $t$ for a period $t_{w2}$ ($\approx 1.5 \times 10^{4}$ sec) at
$T-\Delta T$.  The system was again heated back to $T$ (the
second $T$-shift).  These processes were repeated subsequently.  
In the case of $\Delta T$ = 0.05 K (Fig.~\ref{fig:eighteen}(a)), just after
the first $T$-shift $\chi^{\prime \prime}$ behaves as if the system were
quenched to $T-\Delta T$ directly from 10 K. 
After the second $T$-shift, however, $\chi^{\prime \prime}(\omega,t)$
coincides with a simple extension of $\chi^{\prime \prime}(\omega,t)$
already aged by $t_{w1}$ at $T$.  In the case of $\Delta T$ = 0.20 K
(Figs.~\ref{fig:eighteen}(b) and (c)), $\chi^{\prime\prime}$ undergoes a
drastic change to a value higher than the value just before the change of
$T$ from 3.55 to 3.75 K, when $T$ is changed from 3.75 to 3.55 K. Then the
relaxation of $\chi^{\prime\prime}$ newly occurs (the rejuvenation effect).

\begin{figure}
\includegraphics[width=8.0cm]{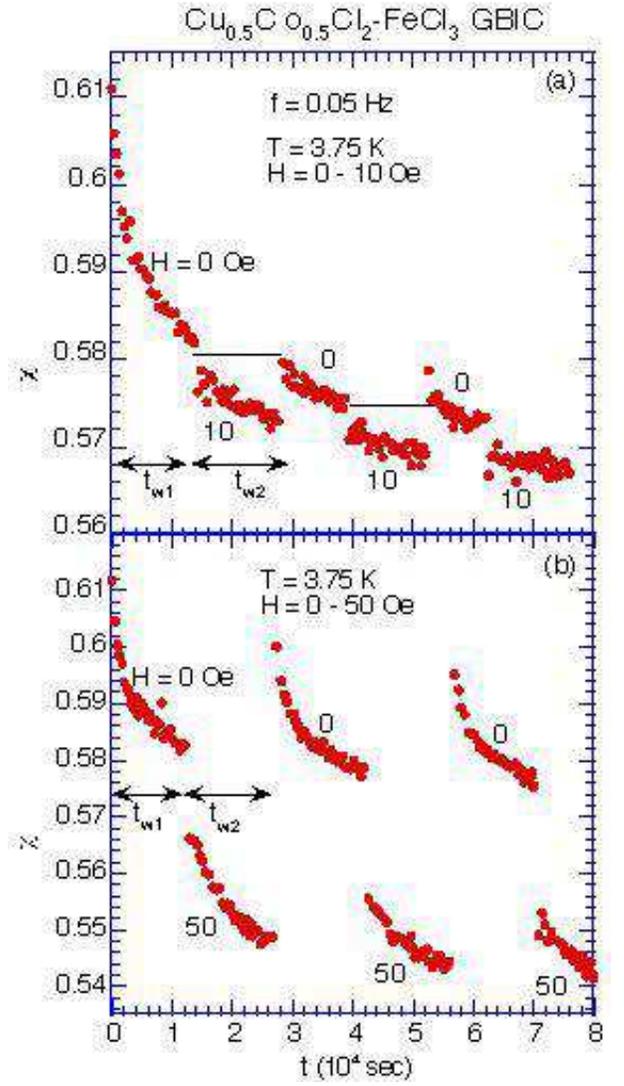}
\caption{\label{fig:ninteen}Relaxation of $\chi^{\prime \prime}(\omega,t)$
during a magnetic field cycling.  $f$ = 0.05 Hz.  $T$ = 3.75 K. (a) $H$ = 0
and 10 Oe.  (b) $H$ = 0 and 50 Oe.}
\end{figure}

Figure \ref{fig:ninteen} shows the $t$ dependence of $\chi^{\prime
\prime}(\omega,t)$ at $T$ = 3.75 K under a positive $H$-cycle, where $f$ =
0.05 Hz and $h$ = 0.1 Oe.  In these measurements, the system was cooled
down to 3.75 K at $H$ = 0.  The relaxation of $\chi^{\prime \prime}$ was
measured as a function of $t$ during a period $t_{w1}$.  A magnetic field
$H$ (= 10 and 50 Oe) was then applied at $t = t_{w1}$.  The relaxation of
$\chi^{\prime \prime}$ was measured as a function of $t$ for a period
$t_{w2}$ at $H$.  The field $H$ was again reduced to zero.  The relaxation
of $\chi^{\prime \prime}$ was measured as a function of $t$ at $H$ = 0. 
These processes were repeated.  In the case of $H$ = 10 Oe
(Fig.~\ref{fig:ninteen}(a)), when the field is off from 10 to 0 Oe,
$\chi^{\prime \prime}$ rises to a value of $\chi^{\prime \prime}$ to which
$\chi^{\prime \prime}$ is supposed to decay during the period $t_{w2}$ at
$H$ = 0.  In the case of $H$ = 50 Oe (Fig.~\ref{fig:ninteen}(b)),
$\chi^{\prime \prime}$ undergoes a drastic change to a value higher than the
value just before the change of $H$ from 0 to 50 Oe, when the field is off
from 50 to 0 Oe.  Then the relaxation of $\chi^{\prime \prime}$ newly
occurs (the rejuvenation effect).  This effect is 
similar to those reported in 
Fe$_{0.5}$Mn$_{0.5}$TiO$_{3}$\cite{Dupuis2001,Jonsson2002} and
CdCr$_{1.7}$In$_{0.3}$S$_{4}$.\cite{Dupuis2001}  

\section{\label{dis}DISCUSSION AND CONCLUSION}
In this paper we study the nature of the slow dynamics of the short-range
Ising SG, Cu$_{0.5}$Co$_{0.5}$Cl$_{2}$-FeCl$_{3}$ GBIC. Using the concepts
of static and dynamic scaling laws, we determine the critical exponents of
this compound.  Our results are as follows: $\alpha = -2.2 \pm 0.5$, $\beta
= 0.36 \pm 0.03$, $\gamma = 3.5 \pm 0.4$, $\nu = 1.4 \pm 0.2$, $\eta = -0.5
\pm 0.2$, $z = 6.6 \pm 1.2$, $\phi = 3.9 \pm 0.4$, $\theta = 0.13 \pm
0.02$, and $\psi = 0.24 \pm 0.02$, where $2-\eta =\gamma / \nu$.  
These critical exponents are compared
with those reported for the Ising SG, Fe$_{0.5}$Mn$_{0.5}$TiO$_{3}$ ($\alpha
= -3$, $\beta = 0.5 \pm 0.2$, $\gamma = 4.0 \pm 0.4$, $\nu = 1.7$, $z =
6.2$, $\phi = 4.5$, $\theta \approx 0.2$, and $\psi = 0.3 -
0.7$)\cite{Gunnarsson1988,Gunnarsson1991,Dupuis2001,Jonsson2002} and
Heisenberg-like SG, CdCr$_{1.7}$In$_{0.3}$S$_{4}$ ($\alpha = -1.9$, $\beta =
0.75 \pm 0.1$, $\gamma = 2.3 \pm 0.4$, $\nu = 1.26 \pm 0.2$, $\phi = 3.1
\pm 0.5$, $z\nu = 7$, $\psi = 1.1$)\cite{Dupuis2001,Vincent1987}.  Our
critical exponents are in excellent agreement with those of amorphous
metallic SG, (Fe$_{0.15}$Ni$_{0.85}$)$_{75}$P$_{16}$B$_{6}$Al$_{3}$ ($\alpha
= -2.2$, $\beta = 0.38$, $\gamma = 3.4$, $\nu = 1.39$, $z =
5.9$).\cite{Lundgren1986,Svedlindh1987} Note that we use the scaling
relation given by $\beta = 2\theta \nu$ to determine $\theta$ of our
system.  This relation is derived in the present work using the idea of
dynamic scaling law for the $T$ and $H$ dependence of $\tau_{-}$.  This
relation is valid for Fe$_{0.5}$Mn$_{0.5}$TiO$_{3}$: the value of $\theta$
(= $\beta/2\nu = 0.15$) is close to the experimental value ($\theta \approx
0.2$).  We note that a relation ($1 = 2\nu \theta$) is derived by Fisher
and Huse\cite{Fisher1988b} in the case of $d \rightarrow d_{l}^{+}$.  Here
$d_{l}^{+}$ is the lowest dimension where $\theta > 0$ and $d_{l}^{+}$ is
close to 2.5.\cite{Campbell2000} If $\beta$ is equal to the mean-field
exponent ($\beta = 1$), then the scaling relation ($\beta = 2\nu \theta$)
coincides with the relation derived by Fisher and Huse.\cite{Fisher1988b}

Our value of $\theta$ is a little smaller than the theoretical values:
$\theta = 0.19 \pm 0.01$ by Bray and Moore,\cite{Bray1987} $\theta = 0.192
\pm 0.001$ by Hartmann,\cite{Hartmann1999} and $\theta = 0.20 \pm 0.03$ by
Komori et al.\cite{Komori1999} from MC simulations on the relaxation of
energy of the 3D Gaussian SG model with nearest neighbor interactions.  Our
value of $\psi$ is a little larger than that of $\theta$, satisfying the
imposed inequality $\theta \leq \psi \leq (d - 1)$.\cite{Fisher1988b} This
condition is also satisfied in the other SG systems:
Fe$_{0.5}$Mn$_{0.5}$TiO$_{3}$ ($\psi = 0.3 - 0.7$)\cite{Dupuis2001} and
CdCr$_{1.7}$In$_{0.3}$S$_{4}$ ($\psi = 1.1$).\cite{Dupuis2001} The value of
$\psi$ is strongly dependent on the spin symmetry ($n$) of the systems such
as Ising ($n = 1$) and Heisenberg ($n = 3$): it decreases as the spin
symmetry $n$ decreases.  This is also true for the critical exponent
$\beta$: $\beta$ decreases with decreasing $n$.\cite{Campbell2000} The
small values of $\psi$ and $\beta$ indicates that our system magnetically
behaves like an ideal 3D short-range Ising SG.

Finally we discuss the $T$ dependence of $b^{\prime \prime}$.  The exponent
$b^{\prime \prime}$ is dependent on $T$.  According to Komori et
al.,\cite{Komori2000} the ratio $\Delta \chi^{\prime \prime}(\omega,
t)/\chi^{\prime}(\omega)$ is described by
\begin{equation} 
\Delta \chi^{\prime \prime}(\omega, t)/\chi^{\prime \prime}(\omega) 
\approx (\omega t)^{-b^{\prime \prime}(T)},
\label{eq:eleven} 
\end{equation} 
where $b^{\prime \prime}(T) = (d - \theta)/z(T)$ and the exponent $1/z(T)$
is linearly dependent on $T$ well below $T_{g}$: $1/z(T) = T/(zT_{g})$. 
This $\omega t$-scaling of $\Delta \chi^{\prime \prime}(\omega,
t)/\chi^{\prime \prime}(\omega)$ is derived on the assumption that the
growth law of the spin-glass correlation length $\xi(t)$ is approximated by
the form $\xi(t) \approx l_{0}(t/t_{0})^{1/z(T)}$, where $l_{0}$ and $t_{0}$
are microscopic length and time scales.  This growth law is different from
that proposed in the scaling picture due to Fisher and
Huse.\cite{Fisher1988a} Using the values of $\theta$ and $b^{\prime
\prime}(T)$, the exponent $1/z(T)$ is calculated as 0.006 at 3.50 K and
0.056 at 3.90 K. Such an increase of $1/z(T)$ with $T$, qualitatively
agrees with the prediction by Komori et al.\cite{Komori2000} If the
expression $1/z(T) = T/(zT_{g})$ is valid at $T = 3.90$ K just below
$T_{g}$, the value of $z$ can be estimated as 18, which is much larger than
our value $z$ (= 6.6).  Experimentally $b^{\prime \prime}(T)$
is nearly equal to zero below $\approx 3$ K ($\approx 0.75 T_{g}$). 
At the present we
give no satisfactory explanation for the cause of the drastic change in
$b^{\prime \prime}$ around 3.5 K in our system.

In conclusion, we have shown that Cu$_{0.5}$Co$_{0.5}$Cl$_{2}$-FeCl$_{3}$
GBIC magnetically behaves like an ideal 3D short-range Ising SG. This compound
undergoes a SG phase transition at $T_{g} = 3.92 \pm 0.11$ K. The dynamic
scaling analysis suggests that this SG transition is destroyed by the
application of $H$ (at least above 100 Oe).  The scaling behavior of the
relaxation time is well described by the scaling picture with the energy
exponent $\theta = 0.13 \pm 0.02$ and the barrier exponent $\psi = 0.24 \pm
0.02$.  The aging obeys the $\omega t$-scaling.
The rejuvenation effect is observed under sufficiently large 
(temperature and magnetic-field) perturbations.

In spite of such a similarity in dynamic behaviors between
Cu$_{0.5}$Co$_{0.5}$Cl$_{2}$-FeCl$_{3}$ GBIC and 3D Ising spin glass, the
frequency dependence of $\chi^{\prime\prime}$ below \textit{T}$_{g}$ are
very different.  The absorption $\chi^{\prime\prime}$ in
Cu$_{0.5}$Co$_{0.5}$Cl$_{2}$-FeCl$_{3}$ GBIC decreases with increasing
frequency, whereas $\chi^{\prime\prime}$ in 3D Ising spin glass increases
with increasing frequency.

\begin{acknowledgments}
We would like to thank H. Suematsu for providing us with single crystal
kish graphite and T. Shima, B. Olson, and M. Johnson for their assistance in
sample preparation and x-ray characterization.  Early work, in particular
for the sample preparation, was supported by NSF DMR 9201656.
\end{acknowledgments}


\begin{references}
\bibitem{Marinari1998}E. Marinari, G. Parisi, and J.J. Ruiz-Lorenzo, in
\textit{Spin Glasses and Random Fields}, edited by A. P. Young (World
Scientific, Singapore, 1998) p.59.  See also references therein.
\bibitem{McMillan1984} W. L. McMillan, J. Phys. C {\bf 17}, 
3179(1984).
\bibitem{Fisher1986} D. S. Fisher and D. A. Huse, Phys. Rev. Lett. 
{\bf 56}, 1601 (1986).
\bibitem{Fisher1988a} D. S. Fisher and D. A. Huse, Phys. Rev. B {\bf
38}, 373 (1988).
\bibitem{Fisher1988b} D.S. Fisher and D.A. Huse, Phys. Rev. B {\bf
38}, 386 (1988).
\bibitem{Bray1987} A.J. Bray and M. A. Moore, Phys. Rev. Lett. {\bf 
58}, 57 (1987).
\bibitem{Bray1988} A. J. Bray, Comments Cond. Mat. Phys. {\bf 14}, 21 
(1988).
\bibitem{Almeida1978} J. R. L. de Almeida and D. J. Thouless, J. 
Phys. A{\bf 11}, 983 (1978).
\bibitem{Mattsson1995} J. Mattsson, T. Jonsson, P. Nordblad, H. A.
Katori, and A. Ito, Phys. Rev. Lett. {\bf 74}, 4305 (1995).
\bibitem{Nordblad1998} P. Nordblad and P. Svedlindh, in \textit{Spin
Glasses and Random Fields}, edited by A. P. Young (World Scientific,
Singapore, 1998) p.1.  See also references therein.
\bibitem{Petit1999} D. Petit, L. Fruchter, and I. A. Campbell, Phys. 
Rev. Lett. {\bf 83}, 5130 (1999).
\bibitem{Suzuki1997}I.S. Suzuki, M. Suzuki, H. Satoh, and T. Enoki,
Solid State Commun. {\bf 104}, 581 (1997).
\bibitem{Suzuki1999a}M. Suzuki and I.S. Suzuki, Phys. Rev. B {\bf
59}, 4221 (1999).
\bibitem{Enoki2003}T. Enoki, M. Suzuki, and M. Endo, in \textit{Graphite Intercalation 
Compounds and Applications} (Oxford Universitry Press, Oxford, 2003) p.236.
\bibitem{Suzuki1998}I. S. Suzuki and M. Suzuki, Solid State Commun. 
{\bf 106}, 513 (1998).
\bibitem{Suzuki1999b}I. S. Suzuki and M. Suzuki, J. Phys. Condensed
Matter {\bf 11}, 521 (1999).
\bibitem{Ito1986}A. Ito, H. Aruga, E. Torikai, M. Kikuchi, Y. Syono,
and H. Takei, Phys. Rev. Lett. {\bf 57}, 483 (1986).
\bibitem{Hansen2002}M.F. Hansen, P. E. J\"{o}nsson, P. Nordblad, and P. Svedlindh, 
J. Phys. Condens. Matter {\bf 14}, 4901 (2002).
\bibitem{Gunnarsson1988}K. Gunnarsson, P. Svedlindh, P. Nordblad, and
L. Lundgren, H. Aruga, and A. Ito, Phys. Rev. Lett. {\bf 61}, 754 
(1988).
\bibitem{Ogielski1985} A.T. Ogielski, Phys. Rev. B {\bf 32}, 7384 
(1985).
\bibitem{Bouchaud2001}J.-P. Bouchaud, V. Dupuis, J. Hammann, and E.
Vincent, Phys. Rev. B {\bf 65}, 024439 (2001).
\bibitem{Svedlindh1989}P. Svedlindh, K. Gunnarsson, P. Nordblad, and
L. Lundgren, H. Aruga, and A. Ito, Phys. Rev. B {\bf 40}, 7162 (1989).
\bibitem{Paulsen1987}C.C. Paulsen, S.J. Williamson, and H. Maletta, Phys. Rev. 
Lett. {\bf 59}, 128 (1987).
\bibitem{Geschwind1990}S. Geschwind, D. A. Huse, and G. E. Devlin, Phys. 
Rev. B {\bf 41}, 4854 (1990).
\bibitem{Gunnarsson1991}K. Gunnarsson, P. Svedlindh, P. Nordblad, L.
Lundgren, H. A. Katori, and A. Ito, Phys. Rev. B {\bf 43}, 8199 
(1991).
\bibitem{Katori1994}H. A. Katori and A. Ito, J. Phys. Soc. Jpn. {\bf
63}, 3122 (1994).
\bibitem{Fischer1991}K.H. Fischer and J.A. Hertz, \textit{Spin
Glasses} (Cambridge University Press, 1991) p.263.
\bibitem{Lamarcq2002}J. Lamarcq, J.-P. Bouchaud, and O. C. Martin,
arXiv: cond-mat/0208100 (2002).
\bibitem{Houdayer1999}J. Houdayer and O. C. Martin, Phys. Rev. Lett. 
{\bf 82}, 4934 (1999).  See also references therein.
\bibitem{Dupuis2001}V. Dupuis, E. Vincent, J.-P. Bouchaud, J. Hammann,
A. Ito, and H. A. Katori, Phys. Rev. B {\bf 64}, 174204 (2001).
\bibitem{Picco2001}M. Picco, F. Ricci-Tersenghi, and F. Ritort, Eur. 
Phys. J. B {\bf 21}, 211 (2001).
\bibitem{Colla2000}E. V. Colla, L. K. Chao, M. B. Weissman, and D. D.
Viehland, Phys. Rev. Lett. {\bf 85}, 3033 (2000).
\bibitem{Jonsson2002}P. E. J\"{o}nsson, H. Yoshino, P. Nordblad, H. A.
Katori, and A. Ito, Phys. Rev. Lett. {\bf 88}, 257204 (2002).
\bibitem{Vincent1987}E. Vincent and J. Hammann, J. Phys. C {\bf 20},
2659 (1987).
\bibitem{Lundgren1986} L. Lundgren, P. Nordblad, and P. Svedlindh,
Phys. Rev. B {\bf 34}, 8164 (1986).
\bibitem{Svedlindh1987}P. Svedlindh, L. Lundgren, P. Nordblad, and
H. S. Chen, Europhys. Lett. {\bf 3}, 243 (1987).
\bibitem{Campbell2000}I. A. Campbell, D. Petit, P. O. Mari, and L. W.
Bernardi, J. Phys. Soc. Jpn. {\bf 69} Suppl. A, 186 (2000).
\bibitem{Hartmann1999}A. K. Hartmann, Phys. Rev. E {\bf 59}, 84 
(1999).
\bibitem{Komori1999}T. Komori, H. Yoshino, and H. Takayama, J. Phys. 
Soc. Jpn. {\bf 68}, 3387 (1999).
\bibitem{Komori2000}T. Komori, H. Yoshino, and H. Takayama, J. Phys. 
Soc. Jpn. {\bf 69}, 1192 (2000).
\end{references}
\end{document}